\documentclass[11pt,draftclsnofoot,onecolumn,peerreviewca]{IEEEtran}

\usepackage[cmex10]{amsmath}
\usepackage{graphicx}
\usepackage{amssymb}
\usepackage{cite}

\newtheorem{definition}{Definition}
\newtheorem{proposition}{Proposition}
\newtheorem{remark}{Remark}
\newtheorem{lemma}{Lemma}

\begin{document}

\title{Cooperative Jamming for Secure Communications in MIMO Relay Networks}

\author{Jing~Huang and A.~Lee~Swindlehurst \thanks{This work was
    supported by the U.S. Army Research Office under the
    Multi-University Research Initiative (MURI) grant
    W911NF-07-1-0318. Part of the results in this work have been
    presented in IEEE GLOBECOM 2010 and ICASSP 2011.}  \thanks{The
    authors are with the Dept. of Electrical Engineering \& Computer
    Science, University of California, Irvine, CA 92697-2625,
    USA. (Email: \{jing.huang; swindle\}@uci.edu). } }

\begin{titlepage}
\maketitle
\begin{abstract}
Secure communications can be impeded by eavesdroppers in conventional relay systems. This paper proposes cooperative jamming strategies for two-hop relay networks where the eavesdropper can wiretap the relay channels in both hops. In these approaches, the normally inactive nodes in the relay network can be used as cooperative jamming sources to confuse the eavesdropper. Linear precoding schemes are investigated for two scenarios where single or multiple data streams are transmitted via a decode-and-forward (DF) relay, under the assumption that global channel state information (CSI) is available. For the case of single data stream transmission, we derive closed-form jamming beamformers and the corresponding optimal power allocation. Generalized singular value decomposition (GSVD)-based secure relaying schemes are proposed for the transmission of multiple data streams. The optimal power allocation is found for the GSVD relaying scheme via geometric programming. Based on this result, a GSVD-based cooperative jamming scheme is proposed that shows significant improvement in terms of secrecy rate compared to the approach without jamming. Furthermore, the case involving an eavesdropper with unknown CSI is also investigated in this paper. Simulation results show that the secrecy rate is dramatically increased when inactive nodes in the relay network participate in cooperative jamming.   
\end{abstract}


\end{titlepage}

\section{Introduction}

Security is an important concern in wireless networks due to their
vulnerability to eavesdropping. Traditionally, security is viewed as
an issue addressed above the physical (PHY) layer, and all widely used
cryptographic protocols are designed and implemented assuming the
physical layer has already been established and provides an error-free
link \cite{Bloch_Wireless08}. However, higher-layer key distribution
and management may be difficult to implement and vulnerable to attack
in complex environments such as ad-hoc or relay networks, in which
transceivers may join or leave randomly
\cite{Schneier_Cryptographic98,Debbah_Mobile08}. Therefore, there has
recently been considerable interest in physical layer security, which
explores the characteristics of the wireless channel to improve
wireless transmission security.

The theoretical basis of this area was laid by Wyner, who introduced
the wiretap channel and demonstrated that when the eavesdropper's
channel is a degraded version of the channel of the legitimate
receiver, the transmitter can send secret messages to the destination
while keeping the eavesdropper from learning anything about the
message \cite{Wyner_wire-tap75}.  The notion of secrecy capacity was
introduced and defined as the maximum achievable transmission rate of
confidential information from the source to its intended
receiver. Later, Csisz\'{a}r and K\"{o}rner generalized Wyner's
approach by considering the transmission of secret messages over
broadcast channels \cite{Csiszar_Broadcast78}. Recently, considerable
research has examined secrecy in wiretap channels with multiple
antennas \cite{Parada_Secrecy05, Khisti_Secure10,
  Shafiee_Achievable07,Liu_MMSE09,
  Oggier_secrecy08,Liu_Note09,Li_Secret07,Khisti_Gaussian07,Goel_Guaranteeing08}. In
particular, the secrecy capacity of the multiple-input multiple-output
(MIMO) wiretap channel has been fully characterized in
 \cite{Oggier_secrecy08,Liu_Note09}. With the additional
degrees of freedom provided by multi-antenna systems, transmitters can
generate artificial noise to degrade the channel condition of the
eavesdropper while maintaining little interference to legitimate users
\cite{Khisti_Gaussian07,Goel_Guaranteeing08,Swindlehurst_Fixed09,Huang_Robust11}.

As a natural extension, approaches for physical layer security have
also been investigated in cooperative relaying networks
\cite{Oohama_Capacity07,Tekin_General08,He_Two-Hop09,Lai_Relay--Eavesdropper08,Dong_Improving10,Zhang_Collaborative10}. In
these cases, relays or even destinations can be used as helpers to
provide jamming signals to confuse the eavesdropper. This approach is
often referred to as cooperative jamming. In
\cite{Lai_Relay--Eavesdropper08}, a noise-forwarding strategy is
introduced for a four-terminal relay-eavesdropper channel where the
full-duplex relay sends codewords independent of the secret message to
confuse the eavesdropper. A two-stage cooperative jamming protocol is
investigated in \cite{Goel_Guaranteeing08}, where multiple relay nodes
act as an extension of the single-antenna source node. In this work,
the ``relays'' only play the role of a helper and do not relay the
information signals. In \cite{Dong_Improving10}, three cooperative
schemes are proposed for a single-antenna relay network, and the
corresponding relay weights and power allocation strategy are derived
to enhance the secrecy for the second hop.  An optimal beamforming
design for decode-and-forward (DF) relays is investigated in
\cite{Zhang_Collaborative10}, but only the scenario where the
eavesdropper wiretaps just the link between the relay and destination
is considered.

Unlike the aforementioned work, this paper proposes cooperative jamming strategies for a half-duplex two-hop wireless MIMO relay system in which the eavesdropper can wiretap the channels during both transmission phases. Cases involving both single and multiple data stream transmissions are investigated. Due to the lack of ``outer'' helpers, the source, relay and destination must rely on themselves for jamming support. This approach guarantees that the eavesdropper is jammed whether it is
close to the source or the destination. In the proposed cooperative jamming strategies, the source and the destination nodes act as temporary helpers to transmit jamming signals during the transmission phases in which they are normally inactive. We define two types of cooperative jamming schemes, \emph{full cooperative jamming} (FCJ) and \emph{partial cooperative jamming} (PCJ), depending on whether or not both the transmitter and the temporary helper transmit jamming signals at the same time.

We focus on the design of linear precoding schemes throughout the paper, and begin with a simple scenario where the relay has only a single antenna. In this case, we investigate the joint design of the jamming beamformer and the power allocation for two optimization problems: (1) maximizing the secrecy rate with certain power constraints, and (2) minimizing the transmit power with a fixed target secrecy rate. Since a joint optimization of the beamformers and power allocation is in general intractable even if global CSI is available, we use a suboptimal zero-forcing constraint that the jamming and information signals lie in orthogonal subspaces when received by the legitimate nodes, and we derive  closed-form expressions for the jamming beamformers. Based on these results, we find the optimal solution for the power allocation by utilizing the method of geometric programming (GP). Then we expand the scope to study the scenario where all nodes have multiple antennas, and multiple data streams are transmitted via the relay. A generalized singular value decomposition (GSVD)-based cooperative jamming scheme is proposed and the corresponding power allocation strategy is discussed. Unlike the single data stream case that uses a zero-forcing constraint, the cooperative GSVD-based jamming method will not in general produce jamming signals that are orthogonal to the desired signal.

Another important consideration is the availability of the eavesdropper's CSI. If the CSI of the eavesdropper is known, (for example, if the eavesdropper is another active user in the wireless network), the transmitter can optimize its beamformer to enhance the information transmission to intended nodes while suppressing or even eliminating the leakage to eavesdroppers. However, in some cases (\textit{e.g.}, passive eavesdroppers), it is impractical to assume known CSI for the eavesdroppers. Since the secrecy rate can not be optimized without knowledge of the eavesdropper's CSI, we will follow the approach of \cite{Swindlehurst_Fixed09,Wang_Cooperative09,Mukherjee_Fixed-rate09,Mukherjee_Robust11}, where the transmitter first allocates part of its resources to guarantee a fixed target rate, and then uses the remaining resources to jam the eavesdropper.

The organization of the paper is as follows. Section \ref{sec:sm} describes the system model considered throughout the paper. In Section \ref{sec:gsvd}, the cooperative jamming schemes, including the jamming beamformer design and power allocation, is investigated when the eavesdropper's CSI is known. Both single and multiple data stream transmissions are considered in this section. Secure relaying under the assumption of unknown eavesdropper's CSI is studied in Section \ref{sec:fixrate}. The performance of the proposed cooperative jamming schemes are discussed in Section \ref{sec:nr}, and conclusions are drawn in Section \ref{sec:con}.

The following notation is used in the paper: $\mathbb{E} \{\cdot\}$ denotes expectation, $(\cdot)^T$ the matrix transpose and $(\cdot)^H$ the Hermitian transpose. $||\cdot||$ represents the Euclidean norm, $|\cdot|$ is the absolute value, $[x]^+$ denotes $\max\{x,0\}$, $\textrm{tr}(\cdot)$ is the trace operator, $\mathcal{N}(\cdot)$ represents the null space, and $\mathbf{I}$ is an identity matrix of appropriate dimension. 


\section{System Model} \label{sec:sm} 
\begin{figure}[ht]
\centering
\includegraphics[width=0.55\textwidth]{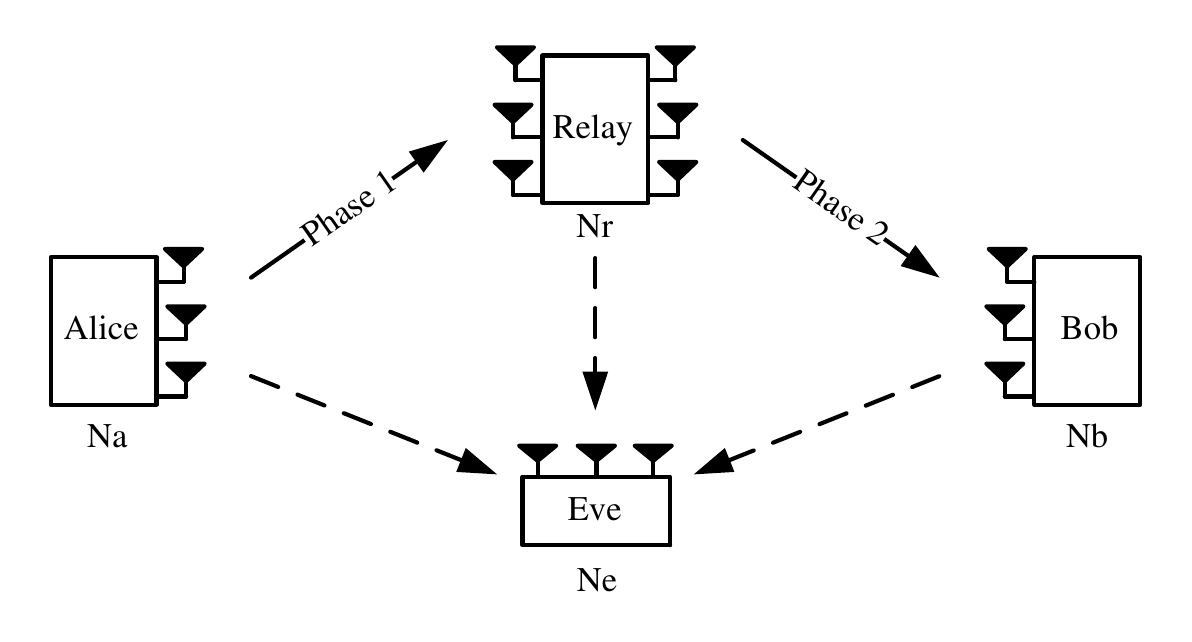}
\caption{\label{fig:sys_mod} Relay scenario.}
\end{figure}

We consider a two-phase
four-terminal relay system composed of a source (Alice), a destination
(Bob), a DF relay node and an eavesdropper (Eve), as shown in
Fig.~\ref{fig:sys_mod}.  The message from Alice is uniformly
distributed over the message set $\mathcal{W}=\{1,\cdots,2^{nR} \}$,
where $R$ denotes the source rate in bits per channel use. The
confidential message is randomly mapped to a length-$n$ source
codeword $z_a^n \in \mathcal{Z}_a^n$ and the Relay encoder maps its
received signal to codeword $z_r^n \in \mathcal{Z}_r^n$, where
$\mathcal{Z}_a^n$ and $\mathcal{Z}_r^n$ are length-n input alphabets.

All nodes are assumed to be half-duplex, \textit{i.e.} a two-hop time division multiple access system is considered. Alice transmits in the first phase while the relay listens, and relay transmits in the second phase.  We assume there is no direct communication link between Alice and Bob, except perhaps for some low-rate control or channel state information, and thus Alice and Bob must rely on two-phase transmissions through the relay.  This is a reasonable assumption in the type of scenarios where relaying is used, where direct high-rate communication is too ``expensive'' in terms of the given power constraints, but low-rate control information can still be exchanged \cite{He_Two-Hop09}.  When Alice transmits a jamming signal, however, its impact on Bob's received signal must be taken into account.   All nodes in general have multiple antennas. The number of antennas possessed by Alice, Bob, the Relay and Eve are denoted by $N_a$, $N_b$, $N_r$ and $N_e$, respectively. In part of the paper, we will explicitly consider scenarios where the Relay has only a single antenna. We restrict attention to scenarios where all nodes (including the eavesdropper) employ linear precoding and receive beamforming.

\subsection{Relay Transmission } \label{sec:repcod}
In the first phase, Alice transmits the information signal to the Relay. Both the Relay and Eve will receive the signal as
\begin{align}
\mathbf{y}_r&=\mathbf{H}_{ar} \mathbf{T}_a \mathbf{z}_a + \mathbf{n}_r \\
\mathbf{y}_{e1}&=\mathbf{H}_{ae} \mathbf{T}_a \mathbf{z}_a + \mathbf{n}_{e1}
\end{align}
where $\mathbf{z}_a$ is the information signal vector transmitted by Alice,
$\mathbf{T}_a \in \mathbb{C}^{N_a \times k}$ $(1 \le k \le s)$
is the transmit beamformer used by Alice, and we assume $m=rank\{\mathbf{H}_{ar}\}$, $n=rank\{\mathbf{H}_{rb}\}$, $s=\min(m,n)$ and $k$ represents the number of data streams to be transmitted. The terms $\mathbf{n}_r$ and
$\mathbf{n}_{e1}$ represent naturally occurring noise at the Relay and Eve, respectively. For simplicity, we assume that the noise vectors at all nodes are Gaussian with covariance $\sigma^2 \mathbf{I}$. In general, $\mathbf{H}_{ij}$
($\mathbf{h}_{ij}$) represents the channel matrix from node $i$ to $j$, with $i,j
\in \{a,b,e,r\}$ denoting which of the four terminals is involved. These channel matrices are fixed over both hops. The signal received by Bob and Eve in the second transmission phase can be expressed as
\begin{align}
\mathbf{y}_b&=\mathbf{H}_{rb} \mathbf{T}_r \mathbf{z}_r + \mathbf{n}_b\\
\mathbf{y}_{e2}&=\mathbf{H}_{re} \mathbf{T}_r \mathbf{z}_r +\mathbf{n}_{e2}
\end{align}
where $\mathbf{z}_r$ is the signal vector transmitted by the Relay,
$\mathbf{T}_r \in \mathbb{C}^{N_r \times k}$ is the transmit beamformer used by the Relay, and
$\mathbf{n}_b$, $\mathbf{n}_{e2}$ represent the noise vectors at Bob and Eve. There is a transmit power constraint $P$ on both phases, \textit{i.e.}, $\mathbb{E}\{\mathbf{z}_a^H \mathbf{z}_a\} \le P$ and $\mathbb{E}\{\mathbf{z}_r^H \mathbf{z}_r\} \le P$. We assume a repetition-coding scheme, where $\mathbf{z}_r$ is simply a scaled version of $\mathbf{z}_a$. In particular, we assume $\mathbf{z}_a=\mathbf{D}_a \mathbf{z}$ and $\mathbf{z}_r=\mathbf{D}_r \mathbf{z}$, where $\mathbb{E}\{\mathbf{z} \mathbf{z}^H\}=\mathbf{I}$ and $\mathbf{D}_a$, $\mathbf{D}_r$ are diagonal power loading matrices that ensure the power constraints are met.

\subsection{Cooperative Jamming} \label{sec:cjmod}
In the most general case, the signals transmitted by Alice in the first phase may contain both information and jamming signals, and Bob may also transmit jamming signals at the same time. Thus the signals received by the Relay and Eve in the first phase will be given by
\begin{align} \label{eq:cj_r_model}
\mathbf{y}_r&=\mathbf{H}_{ar} (\mathbf{T}_{a} \mathbf{z}_{a} + \mathbf{T}_{a}^\prime \mathbf{z}_{a}^\prime) + \mathbf{H}_{br} \mathbf{T}_{b}^\prime  \mathbf{z}_{b}^\prime + \mathbf{n}_r \\
\mathbf{y}_{e1}&=\mathbf{H}_{ae} (\mathbf{T}_{a} \mathbf{z}_{a} + \mathbf{T}_{a}^\prime \mathbf{z}_{a}^\prime) + \mathbf{H}_{be} \mathbf{T}_{b}^\prime  \mathbf{z}_{b}^\prime + \mathbf{n}_{e1}
\end{align}
where $\mathbf{z}_{a}^\prime$ and $\mathbf{z}_{b}^\prime$ are jamming signal vectors transmitted by Alice and Bob, respectively, and $\mathbf{T}_{a}^\prime$ and $\mathbf{T}_{b}^\prime$ are the corresponding transmit beamformers.
In this paper, $\mathbf{T}_{a}^\prime$ and $\mathbf{T}_{b}^\prime$ could be chosen to project the jamming signals on the subspace orthogonal to the information signals, or they could allow a small amount of interference leakage to the legitimate receiver while producing more interference power at Eve, as will be discussed when the GSVD-based transmission strategy is used. We refer to the case where both $\mathbf{z}_{a}^\prime \ne \mathbf{0}$ and $\mathbf{z}_{b}^\prime \ne \mathbf{0}$ as \emph{full cooperative jamming} (FCJ). If either of them is zero, we refer to it as \emph{partial cooperative jamming} (PCJ). FCJ will not be considered in the scenario where Eve's CSI is known, since in this case splitting the power between data and jamming signals at Alice is known to be suboptimal. However, when Eve's CSI is known, we will still study the PCJ scheme where Bob uses part of the global transmit power to produce jamming signals. When Eve's CSI is not available, FCJ should be used, as will be discussed in Section \ref{sec:fixrate}.

In phase 2, the signals received by Bob and Eve are given by
\begin{align}\label{eq:cj_b_model}
\mathbf{y}_b&=\mathbf{H}_{rb} (\mathbf{T}_{r} \mathbf{z}_{r} + \mathbf{T}_{r}^\prime \mathbf{z}_{r}^\prime) + \mathbf{H}_{ab} \mathbf{T}_{a2}^\prime  \mathbf{z}_{a2}^\prime + \mathbf{n}_b \\
\mathbf{y}_{e2}&=\mathbf{H}_{re} (\mathbf{T}_{r} \mathbf{z}_{r} + \mathbf{T}_{r}^\prime \mathbf{z}_{r}^\prime) + \mathbf{H}_{ae} \mathbf{T}_{a2}^\prime  \mathbf{z}_{a2}^\prime + \mathbf{n}_{e2} \label{eq:cj_e2_model}
\end{align}
where $\mathbf{z}_{r}$ is the information signal vector of the Relay
with transmit beamformer $\mathbf{T}_r$, $\mathbf{z}_{r}^\prime$ and
$\mathbf{z}_{a2}^\prime$ are jamming signal vectors transmitted by the
Relay and Alice, respectively, and $\mathbf{T}_{r}^\prime$ and
$\mathbf{T}_{a2}^\prime$ are their corresponding transmit
beamformers.   Note that, although there is no direct link for the information
signal, Bob still sees the jamming signal from Alice.  For a global power constraint, we have
\begin{align}
\mathbb{E}\{\mathbf{z}_{a}^H \mathbf{z}_{a} + \mathbf{z}_{a}^{\prime H} \mathbf{z}_{a}^\prime + \mathbf{z}_{b}^{\prime H} \mathbf{z}_{b}^\prime\} \le P \qquad
\mathbb{E}\{\mathbf{z}_r^H \mathbf{z}_r + \mathbf{z}_r^{\prime H} \mathbf{z}_r^\prime + \mathbf{z}_{a2}^{\prime H} \mathbf{z}_{a2}^\prime\} \le P. \notag
\end{align}
We will also investigate scenarios with individual power constraints, \textit{i.e.}
$\mathbb{E}\{\mathbf{z}_{a}^H \mathbf{z}_{a} + \mathbf{z}_{a}^{\prime H} \mathbf{z}_{a}^\prime \} \le P_{a}$,
$\mathbb{E}\{\mathbf{z}_{b}^{\prime H} \mathbf{z}_{b}^\prime\} \le P_b$,
$\mathbb{E}\{\mathbf{z}_r^H \mathbf{z}_r + \mathbf{z}_r^{\prime H} \mathbf{z}_r^\prime \} \le P_r$, and
$\mathbb{E}\{ \mathbf{z}_{a2}^{\prime H} \mathbf{z}_{a2}^\prime\} \le P_a$.

\subsection{Performance Metric}
MIMO wiretap channels have been extensively analyzed in recent work,
and the achievable secrecy rate has been shown to be
\cite{Oggier_secrecy08,Liu_Note09}
\begin{equation} \label{eq:sr}
R_s = \max[I_d - I_e]^+
\end{equation}
where $I_d$ is the mutual information from the source to the
destination, $I_e$ is the mutual information from the source to the
eavesdropper, and the maximum is taken over all possible input
covariance matrices.  For the half-duplex two-hop relay channel, the
achievable secrecy rate was found in \cite{Yuksel_Secure07} to satisfy
the same expression as in \eqref{eq:sr}, where amplify-and-forward,
decode-and-forward, and compress-and-forward relaying modes were all
investigated. Eq.~\eqref{eq:sr} was also used as a performance metric
to evaluate cooperative jamming schemes for half-duplex relay networks
in \cite{Dong_Improving10}.  In general, to obtain the maximum secrecy
rate, one must construct an optimal coding scheme, although
potentially suboptimal Gaussian codebooks are assumed in
\cite{Shafiee_Achievable07,Dong_Improving10,Yuksel_Secure07}.  In
Section~\ref{sec:gsvd}, we will follow the convention adopted in
\cite{Yuksel_Secure07,Dong_Improving10} and use~(\ref{eq:sr}) as our
metric for evaluating the achievable secrecy rate, assuming Gaussian
inputs.  Note that~\eqref{eq:sr} was shown to be valid for both
independent and repetition codebooks \cite{Yuksel_Secure07}, although
we will only focus on repetition coding (\textit{e.g.}
\cite{Laneman_Distributed03,Azarian_achievable05}) at the relay since
independent codebooks are expected to result in smaller secrecy rates
when the encoding schemes and relay protocols are public information
\cite{Yuksel_Secure07}.

The discussion above applies to the cases where the eavesdropper's CSI
is known or at least partially known (\textit{e.g.} the case where only
statistical channel knowledge is available and ergodic secrecy rate is
studied \cite{Shafiee_Achievable07,Li_Ergodic11}). However,
when the eavesdropper's CSI is completely unavailable, \eqref{eq:sr}
may not represent an achievable secrecy rate.  Some recent progress
has been made on finding expressions for the achievable secrecy rate
in certain scenarios where the eavesdropper's CSI is completely
unknown \cite{He_MIMO10}, but the derivation of such an expression for
the relay network considered here is still an open
problem. Nonetheless, the difference in the mutual information between
the desired receiver and the eavesdropper is still a valid metric for
evaluating the relative security of competing physical layer
approaches.  While the transmission parameters cannot be chosen to
optimize~\eqref{eq:sr} when the eavesdropper's CSI is unknown, the
approach of
\cite{Swindlehurst_Fixed09,Mukherjee_Fixed-rate09,Mukherjee_Robust11}
can be followed in which attention is restricted to obtaining a
certain desired QoS for the legitimate receiver, and then finding a
robust strategy for using the remaining resources to jam potential
eavesdroppers.  This is the approach adopted in
Section~\ref{sec:fixrate}, with~\eqref{eq:sr} as the performance
metric.


\section{Secure relaying with known ECSI} \label{sec:gsvd}
In this section, we assume that Eve's CSI (ECSI) is available to the
relay network. We will begin with the simple case where the Relay is
equipped with only a single antenna, then a more complicated scenario
with a MIMO relay will be investigated.

\subsection{Single data stream relaying} \label{sec:single_data}
We begin by assuming a single-antenna DF relay ($N_r=1$), where only
one data stream can be transmitted via the Relay. Under the PCJ
approach, the signals received in each phase can be expressed as
\begin{equation}
y_r=\mathbf{h}_{ar} \mathbf{t}_{a} z_a
+  \mathbf{h}_{br} \mathbf{T}_b^\prime \mathbf{z}_b^\prime
+ n_r
\end{equation}
\begin{equation} \label{eq:single1}
\mathbf{y}_{e1}=\mathbf{H}_{ae} \mathbf{t}_{a} z_a
+  \mathbf{H}_{be} \mathbf{T}_b^\prime \mathbf{z}_b^\prime
+ \mathbf{n}_{e1}
\end{equation}
and
\begin{equation} \label{eq:sm1}
\mathbf{y}_b= \mathbf{h}_{rb}  z_r
+ \mathbf{H}_{ab} \mathbf{T}_a^\prime \mathbf{z}_a^\prime+ \mathbf{n}_b
\end{equation}
\begin{equation} \label{eq:single2}
\mathbf{y}_{e2}= \mathbf{h}_{re} z_r
+  \mathbf{H}_{ae} \mathbf{T}_a^\prime \mathbf{z}_a^\prime
+ \mathbf{n}_{e2}
\end{equation}
where $\mathbb{E}\{z_{a}^H z_{a} \} = p_{a}$, $\mathbb{E}\{\mathbf{z}_{b}^{\prime H} \mathbf{z}_{b}^\prime\} = p_b$, $\mathbb{E}\{z_r^H z_r \} = p_r$ and $\mathbb{E}\{ \mathbf{z}_{a}^{\prime H} \mathbf{z}_{a}^\prime\} = p_{a2}$. This is the PCJ form of (\ref{eq:cj_r_model})-(\ref{eq:cj_e2_model}) with $\mathbf{z}_{a}^\prime = \mathbf{0}$ and $\mathbf{z}_r^\prime = \mathbf{0}$. Since $N_r=1$ in this case, we can design $\mathbf{T}_b^\prime$ such that the jamming signals are
completely nulled at the Relay, \textit{i.e.}, $\mathbf{h}_{br}
\mathbf{T}_b^\prime = \mathbf{0}$. For the transmit beamformer $\mathbf{t}_a$ in the first phase, we choose
the generalized eigenvector of the pencil $(\mathbf{I}+ \frac{p_a}{\sigma^2} \mathbf{h}_{ar}^H
\mathbf{h}_{ar},\mathbf{I}+ \frac{p_a}{\sigma^2}\mathbf{H}_{ae}^H \mathbf{H}_{ae})$ with the largest
generalized eigenvalue, which achieves the secrecy capacity for the
single-hop MISO wiretap channel\cite{Khisti_Gaussian07}. For the second phase, we design $\mathbf{T}_a^\prime$ such that $\mathbf{H}_{ab} \mathbf{T}_a^\prime$ is orthogonal to the one-dimensional signal subspace span$\{\mathbf{h}_{rb}\}$, so that the jamming does not impact Bob's reception of the information signal.

\subsubsection{Maximum secrecy rate with power constraints}
Next, we will discuss the design of the jamming beamformers
and power allocation for maximizing the secrecy rate under both global
power constraints ($p_{a}+p_b \le P$ in the first phase and $p_r+p_{a2} \le P$
in the second phase) and individual power constraints. For a two-hop DF-based relay channel, the mutual information between Alice and Bob through the relay
link can be written as \cite{Laneman_Cooperative04}
\begin{equation}
I_d =\cfrac{1}{2}~ \min\{\log_2(1+\gamma_{ar}),\log_2(1+\gamma_{rb})\}
\end{equation}
where $\frac{1}{2}$ appears because the relay transmission is divided
into two stages, and $\gamma_{ij}$ is the SINR at node $j$ for the
signal from node $i$. Eve receives data during both phases,
and her mutual information is
\begin{equation}
I_e =\cfrac{1}{2}~ \min\{\log_2(1+\gamma_{ar}),\log_2(1+\gamma_{ae}+\gamma_{re}) \}.
\end{equation}
Thus, the secrecy rate can be expressed as
\begin{equation} \label{eq:seccons}
R_s = \left \{
\begin{array}{ll}
\frac{1}{2}~ \log_2 \frac{\min\{1+\gamma_{ar},1+\gamma_{rb}\}}{(1+\gamma_{ae}+\gamma_{re})}, & \gamma_{ae}+\gamma_{re} \le \gamma_{ar} < \gamma_{rb} \textrm{ or } \gamma_{ar} \ge \max\{\gamma_{rb}, \gamma_{ae}+\gamma_{re} \} \\
0, & \textrm{otherwise}.
\end{array} \right.
\end{equation}
Since the rate of the relay link is limited by the
SINR of the inferior phase, for a single data stream the transmit power
for Alice and the Relay should be adjusted such that
$\gamma_{ar}=\gamma_{rb}$ for power efficiency. Thus $R_s = \frac{1}{2}~ \log_2 \frac{(1+\gamma_{ar})}{(1+\gamma_{ae}+\gamma_{re})}$ will be used as the objective function in the remainder of this section, as a result of the power adjustment.

We assume Eve uses beamformers
$\mathbf{w}_{e1}$ and $\mathbf{w}_{e2}$ to receive the signals from Alice
and the Relay in the first and second phases, respectively:
\begin{align}
\mathbf{w}_{e1}^H \mathbf{y}_{e1} &=
\mathbf{w}_{e1}^H(\mathbf{H}_{ae} \mathbf{t}_{a} z_a +
\mathbf{H}_{be} \mathbf{T}_b^\prime \mathbf{z}_b^\prime + \mathbf{n}_{e1}) \\
\mathbf{w}_{e2}^H \mathbf{y}_{e2} &=
\mathbf{w}_{e1}^H(\mathbf{h}_{re} z_r
+  \mathbf{H}_{ae} \mathbf{T}_a^\prime \mathbf{z}_a^\prime
+ \mathbf{n}_{e2}),
\end{align}
and we assume that Eve can compute the beamformers which yield the best SINR,
\begin{align}
\mathbf{w}_{e1} &= (\mathbf{H}_{be} \mathbf{T}_b^\prime \mathbf{Q}_{zb^\prime}
\mathbf{T}_b^{\prime H} \mathbf{H}_{be}^H + \sigma^2 \mathbf{I})^{-1} \mathbf{H}_{ae}
\mathbf{t}_{a} \\
\mathbf{w}_{e2} &= (\mathbf{H}_{ae} \mathbf{T}_a^\prime \mathbf{Q}_{za^\prime} \mathbf{T}_a^{\prime H} \mathbf{H}_{ae}^H + \sigma^2 \mathbf{I})^{-1} \mathbf{h}_{re}
\end{align}
where $\mathbf{Q}_{zb^\prime}=\mathbb{E}\{ \mathbf{z}_b^\prime \mathbf{z}_b^{\prime H}\}$ and $\mathbf{Q}_{za^\prime}=\mathbb{E}\{ \mathbf{z}_a^\prime \mathbf{z}_a^{\prime H} \}$. With the above assumptions, the secrecy rate can be written
as
\begin{equation} \label{eq:rs1}
R_s = \cfrac{1}{2}~ \log_2 \cfrac{(1+\gamma_{ar})}{(1+\gamma_{ae}+\gamma_{re})}
\end{equation}
where
\begin{align}
\gamma_{ar} &= \frac{p_{a}}{\sigma^2} |\mathbf{h}_{ar} \mathbf{t}_{a} |^2 \\
\gamma_{ae} &= p_{a} \mathbf{t}_{a}^H \mathbf{H}_{ae}^H (\mathbf{H}_{be} \mathbf{T}_b^\prime \mathbf{Q}_{zb^\prime} \mathbf{T}_b^{\prime H} \mathbf{H}_{be}^H + \sigma^2 \mathbf{I})^{-1} \mathbf{H}_{ae} \mathbf{t}_{a} \label{eq:gmae} \\
\gamma_{re} &= p_r \mathbf{h}_{re}^H (\mathbf{H}_{ae} \mathbf{T}_a^\prime \mathbf{Q}_{za^\prime} \mathbf{T}_a^{\prime H} \mathbf{H}_{ae}^H + \sigma^2 \mathbf{I})^{-1} \mathbf{h}_{re},
\end{align}
and we aim to find the joint optimal solution for the jamming beamformers
$\mathbf{T}_a^\prime $, $\mathbf{T}_b^\prime$, the covariance matrices $\mathbf{Q}_{zb^\prime}$, $\mathbf{Q}_{za^\prime}$, and the transmit power vector
$\mathbf{p}=[p_{a},p_r,p_{a2},p_b]^T$ in order to maximize
the secrecy rate $R_s$.

We will first consider optimizing the jamming beamformers and covariance matrices. For $\mathbf{T}_b^\prime$ and $\mathbf{Q}_{zb^\prime}$, the problem of minimizing the SINR at Eve $\gamma_{ae}$ can be written as
\begin{subequations} \label{eq:qzb}
\begin{align}
\min_{\mathbf{Q}_{zb^\prime} \succeq 0, \mathbf{T}_b^\prime}& \quad  \mathbf{t}_{a}^H \mathbf{H}_{ae}^H (\mathbf{H}_{be} \mathbf{T}_b^\prime \mathbf{Q}_{zb^\prime} \mathbf{T}_b^{\prime H} \mathbf{H}_{be}^H + \sigma^2 \mathbf{I})^{-1} \mathbf{H}_{ae} \mathbf{t}_{a} \label{eq:qzba} \\
 \textrm{s.t.}& \quad  \textrm{tr}(\mathbf{Q}_{zb^\prime}) \le p_b, \ \mathbf{h}_{br} \mathbf{T}_b^\prime = \mathbf{0} .
\end{align}
\end{subequations}
Although problem \eqref{eq:qzb} can be formulated as a semidefinite program (SDP) that can be solved efficiently (see Appendix \ref{sec:appsdp}), we can not directly obtain an analytical solution that is useful for optimizing the global power allocation. Therefore, we will make use of the following lemma, a proof of which is provided in Appendix \ref{sec:apprk1}.
\begin{lemma} \label{rk1}
The covariance matrix $\mathbf{Q}_{zb^\prime}$ that minimizes \eqref{eq:qzba} is rank one.
\end{lemma}

According to Lemma \ref{rk1}, we know that a one-dimensional jamming signal is optimal for the case of single data stream
transmission: $\mathbf{T}_b^\prime = \mathbf{t}_b^\prime$. Under the constraint that $\mathbf{h}_{br} \mathbf{t}_b^\prime
= 0$, and defining $\mathbf{G}_b^\perp$ as an orthonormal basis
for $\mathcal{N}(\mathbf{h}_{br})$, the
jamming beamformer from Bob can be written as $\mathbf{t}_b^\prime =
\mathbf{G}_b^\perp \mathbf{c}_b$, for some unit-length vector
$\mathbf{c}_b$. Eq.~\eqref{eq:gmae} becomes
\begin{align}
&\gamma_{ae} = p_{a} \mathbf{t}_{a}^H \mathbf{H}_{ae}^H (p_b \mathbf{H}_{be} \mathbf{t}_b^\prime \mathbf{t}_b^{\prime H} \mathbf{H}_{be}^H + \sigma^2 \mathbf{I})^{-1} \mathbf{H}_{ae} \mathbf{t}_{a}  \notag \\
						&= \frac{p_{a}}{\sigma^2} \left( \mathbf{t}_{a}^H \mathbf{H}_{ae}^H  \mathbf{H}_{ae} \mathbf{t}_{a}  - \cfrac{ \mathbf{t}_{a}^H \mathbf{H}_{ae}^H \mathbf{H}_{be} \mathbf{t}_b^\prime \mathbf{t}_b^{\prime H} \mathbf{H}_{be}^H \mathbf{H}_{ae} \mathbf{t}_{a} }{ \mathbf{t}_b^{\prime H} (\frac{\sigma^2}{p_b} \mathbf{I} +  \mathbf{H}_{be}^H \mathbf{H}_{be}) \mathbf{t}_b^\prime } \right),  \label{eq16}
\end{align}
where the second equality holds due to the matrix inversion lemma \cite{Golub_Matrix96}. The optimization problem is equivalent to maximizing the second term in (\ref{eq16}), which can be formulated as
\begin{equation} \label{eq:cj21}
\max_{\mathbf{c}_{b}} \quad \frac{\mathbf{c}_b^H \mathbf{a}_b \mathbf{a}_b^H \mathbf{c}_b}{\mathbf{c}_b^H (\frac{\sigma^2}{p_b} \mathbf{I} +  \mathbf{B}_b^H \mathbf{B}_b) \mathbf{c}_b }  \qquad
\mathrm{s.t.} \quad \mathbf{c}_{b}^H \mathbf{c}_{b} = 1
\end{equation}
where $\mathbf{a}_b=\mathbf{G}_b^{\perp H} \mathbf{H}_{be}^H
\mathbf{H}_{ae} \mathbf{t}_a$ and $\mathbf{B}_b = \mathbf{H}_{be}
\mathbf{G}_b^\perp$.  The maximum value of the Rayleigh quotient in
\eqref{eq:cj21} is the largest generalized eigenvalue of the matrix
pencil $(\mathbf{a}_b \mathbf{a}_b^H,\frac{\sigma^2}{p_b} \mathbf{I} +
\mathbf{B}_b^H \mathbf{B}_b)$, and the vector that achieves it is
the corresponding generalized eigenvector \cite{Khisti_Secure10a}. Since $\mathbf{a}_b
\mathbf{a}_b^H$ is rank one, the solution can be written as
\begin{equation} \label{eq14}
\mathbf{t}_b^\prime = \mathbf{G}_b^\perp \frac{(\frac{\sigma^2}{p_b} \mathbf{I} +
\mathbf{B}_b^H \mathbf{B}_b)^{-1} \mathbf{a}_b}{||(\frac{\sigma^2}{p_b}
\mathbf{I} + \mathbf{B}_b^H \mathbf{B}_b)^{-1} \mathbf{a}_b||} \ ,
\end{equation}
and $\gamma_{ae}$ becomes
\begin{equation}
\gamma_{ae} = \frac{p_{a}}{\sigma^2} \left( \mathbf{t}_{a}^H \mathbf{H}_{ae}^H
\mathbf{H}_{ae} \mathbf{t}_{a} - \mathbf{a}_b^H (\frac{\sigma^2}{p_b}
\mathbf{I} + \mathbf{B}_b^H \mathbf{B}_b)^{-1} \mathbf{a}_b \right) .
\end{equation}

Similarly for the second phase, the SINR for Eve is rewritten as
\begin{align}
\gamma_{re} &= p_r \mathbf{h}_{re}^H (p_{a2} \mathbf{H}_{ae} \mathbf{t}_a^\prime \mathbf{t}_a^{\prime H} \mathbf{H}_{ae}^H + \sigma^2 \mathbf{I})^{-1} \mathbf{h}_{re} \notag \\
						&= \frac{p_r}{\sigma^2} \left( \mathbf{h}_{re}^H  \mathbf{h}_{re} - \cfrac{ \mathbf{h}_{re}^H \mathbf{H}_{ae} \mathbf{t}_a^\prime \mathbf{t}_a^{\prime H} \mathbf{H}_{ae}^H \mathbf{h}_{re}}{\mathbf{t}_a^{\prime H} (\frac{\sigma^2}{p_{a2}} \mathbf{I} + \mathbf{H}_{ae}^H \mathbf{H}_{ae}) \mathbf{t}_a^\prime } \right).
\end{align}
Using the same method as in \eqref{eq16}-\eqref{eq14},
Alice's jamming beamformer is given by
\begin{equation} \label{eq18}
\mathbf{t}_a^\prime = \mathbf{G}_a^\perp \frac{(\frac{\sigma^2}{p_a} \mathbf{I} +
\mathbf{B}_a^H \mathbf{B}_a)^{-1} \mathbf{a}_a}{||(\frac{\sigma^2}{p_a}
\mathbf{I} + \mathbf{B}_a^H \mathbf{B}_a)^{-1} \mathbf{a}_a||} \ ,
\end{equation}
where $\mathbf{G}_a^\perp$ is an orthonormal basis for
$\mathcal{N}(\mathbf{h}_{rb}^H \mathbf{H}_{ab})$,
$\mathbf{a}_a=\mathbf{G}_a^{\perp H} \mathbf{H}_{ae}^H
\mathbf{h}_{re}$,$\mathbf{B}_a = \mathbf{H}_{ae} \mathbf{G}_a^\perp$,
and $\gamma_{re}$ becomes
\begin{equation}
\gamma_{re} = \frac{p_r}{\sigma^2} \left( \mathbf{h}_{re}^H \mathbf{h}_{re} -
\mathbf{a}_a^H (\frac{\sigma^2}{p_a}
\mathbf{I} + \mathbf{B}_a^H \mathbf{B}_a)^{-1} \mathbf{a}_a \right) \notag .
\end{equation}

Next we find the power allocation that maximizes the secrecy
rate. Note that the jamming beamformers are not independent of the jamming
power, and thus we need to jointly optimize over both quantities. In general, (\ref{eq:rs1}) is not convex with respect to $\mathbf{p}$, so instead we maximize the following lower bound for $R_s$:
\begin{equation} \label{eq19}
R_s(\mathbf{p}) \ge \cfrac{1}{2}~ \log_2
	\cfrac{\gamma_{ar}}{(1+\gamma_{ae}+\gamma_{re})}
	= \cfrac{1}{2}~ \log_2
	\cfrac{|\mathbf{h}_{ar} \mathbf{t}_{a} |^2}{\sigma^2 g(\mathbf{p})}
\end{equation}
where
\begin{align}
g(\mathbf{p})&=p_{a}^{-1} + \tilde{p_b}^{-1} + p_{a}^{-1} p_r \tilde{p}_{a2}^{-1} \\
\tilde{p}_b^{-1}&=\mathbf{t}_{a}^H \mathbf{H}_{ae}^H
\mathbf{H}_{ae} \mathbf{t}_{a} - \mathbf{a}_b^H (\frac{\sigma^2}{p_b}
\mathbf{I} + \mathbf{B}_b^H \mathbf{B}_b)^{-1} \mathbf{a}_b \label{eq:pbinv} \\
\tilde{p}_{a2}^{-1}&=\mathbf{h}_{re}^H \mathbf{h}_{re} -
\mathbf{a}_a^H (\frac{\sigma^2}{p_a}
\mathbf{I} + \mathbf{B}_a^H \mathbf{B}_a)^{-1} \mathbf{a}_a . \label{eq:painv}
\end{align}
Over the range of practical transmit powers, $\tilde{p_b}$ and
$\tilde{p}_{a2}$ can be accurately approximated as linear functions of
$p_b$ and $p_{a2}$, which we denote by $\tilde{p_b} = c_1 p_b + c_2$
and $\tilde{p}_{a2} = c_3 p_{a2} + c_4$. Note that according to
\eqref{eq:pbinv}, as $p_b$ increases, the second term can only
increase in size, which means $\tilde{p}_b^{-1}$ decreases, and hence
$\tilde{p}_b$ increases, which implies that $c_1$ is positive. As
$p_b$ approaches zero, the second term approaches zero, but the first
term is non-negative, so that implies that $c_2 > 0$. Thus $c_1$ and
$c_2$ are both positive constants. Similarly, we can see that $c_3$
and $c_4$ in~\eqref{eq:painv} are also positive constants.

Using this approximation, the rate maximization problem under a global
power constraint $P$ becomes one of minimizing $g(\mathbf{p})$ in
\eqref{eq19}:
\begin{subequations}
\begin{align} \label{eq:approx_min}
\min_{\mathbf{p}}& \quad p_{a}^{-1} + \tilde{p_b}^{-1} + p_{a}^{-1} p_r \tilde{p}_{a2}^{-1}  \\
\mathrm{s.t.}& \quad p_{a} + c_1^{-1}\tilde{p_b} \le P+c_1^{-1} c_2,
						   \quad  p_r + c_3^{-1}\tilde{p}_{a2} \le P+c_3^{-1} c_4 \label{eq:posy2} \\
						 & \quad p_a |\mathbf{h}_{ar} \mathbf{t}_a|^2 = p_r |\mathbf{h}_{rb}|^2  \label{eq:posy3}
\end{align}
\end{subequations}
where \eqref{eq:posy2} is derived from $p_{a} + p_b \le P$ and $p_r +
p_{a2} \le P$, and \eqref{eq:posy3} is the optimal power adjustment
for the two hops used to guarantee that $\gamma_{ar}=\gamma_{rb}$. The
optimization problem stated above is in the standard form for
Geometric Programming (GP) problems, with~\eqref{eq:approx_min}
and \eqref{eq:posy2} as posynomial and \eqref{eq:posy3} as monomial constraints.
GP problems are a class of non-linear optimization problems that
can be readily turned into convex optimization problems, and hence a
global optimum can be efficiently computed \cite{Boyd_Convex04}. If
individual power constraints are employed, we can also use GP to solve
the following similar optimization problem:
\begin{subequations}
\begin{align}
\min_{\mathbf{p}}& \quad p_{a}^{-1} + \tilde{p_b}^{-1} + p_{a}^{-1} p_r \tilde{p}_{a2}^{-1}  \\
\mathrm{s.t.}& \quad p_{a} \le P_a,
						  \quad p_r \le P_r  \\
						 & \quad c_1^{-1} \tilde{p_b} \le P_b + c_1^{-1} c_2,
						  \quad  c_3^{-1} \tilde{p}_{a2} \le P_a + c_3^{-1} c_4 \\
						 & \quad p_a |\mathbf{h}_{ar} \mathbf{t}_a|^2 = p_r |\mathbf{h}_{rb}|^2.
\end{align}
\end{subequations}
\begin{remark}
As discussed in the beginning of this section, we choose the principal generalized eigenvector of the pencil $(\mathbf{I}+ \frac{p_a}{\sigma^2} \mathbf{h}_{ar}^H
\mathbf{h}_{ar},\mathbf{I}+ \frac{p_a}{\sigma^2}\mathbf{H}_{ae}^H \mathbf{H}_{ae})$ as the information signal transmit beamformer $\mathbf{t}_a$. However, the allocated power $p_a$ is unavailable before the optimization algorithm starts. Therefore, iterations will be needed for computing the beamformers, initialized with $p_a=P$, where $P$ is the maximum transmit power. Based on our numerical experiments, the algorithm usually converges with very few iterations, and introduces little complexity to the overall algorithm.
\end{remark}
\begin{remark}
For the case where $\mathbf{H}_{ae}$ does not have full column rank, \textit{i.e.}, $N_a>N_e$, an alternative would be to choose $\mathbf{t}_a$ to lie in the null space $\mathcal{N}(\mathbf{H}_{ae})$. This beamformer will in general be different from the one we propose, and will result in a solution where Eve will not receive any information signal in the first phase and hence the jamming from Bob is not necessary. However, based on our numerical experiments, the solution we propose yields a larger secrecy rate.  This is mainly due to the fact that although $\mathbf{t}_a$ may allow a small amount of information leakage from Alice to Eve, the rate improvement in the information channel outweighs that of the wiretap channel, given the cooperative jamming support from Bob and the optimized power allocation.
\end{remark}

\subsubsection{Minimum transmit power with fixed secrecy rate}
The problem of minimizing the transmit power under a certain fixed secrecy rate is similar to the problems discussed above. We still choose jamming
beamformers that lie in the subspace orthogonal to the intended
channels. As before, for the first phase we will have $\mathbf{t}_b^\prime=\rho_b
\mathbf{G_b^\perp} \mathbf{e}_b$, where $\rho_b$ is a scalar that maintains
the unit norm of $\mathbf{t}_b^\prime$. We aim to minimize
the norm of $\mathbf{e}_b$ under a fixed target
secrecy rate $R_0$. According to \eqref{eq:cj21} and \eqref{eq19},
the problem can be formulated as
\begin{equation}
\min_{\mathbf{e}_b} \quad \mathbf{e}_{b}^H  \mathbf{e}_{b} \qquad
\mathrm{s.t.} \quad \frac{\mathbf{c}_b^H \mathbf{a}_b \mathbf{a}_b^H \mathbf{c}_b}{\mathbf{c}_b^H (\frac{\sigma^2}{p_b} \mathbf{I} +  \mathbf{B}_b^H \mathbf{B}_b) \mathbf{c}_b } \ge f(R_0,\mathbf{t}_a^\prime)
\end{equation}
where $f(R_0,\mathbf{t}_a^\prime)$ is a function of $R_0$ and
$\mathbf{t}_a^\prime$ independent of $\mathbf{t}_b^\prime$. The solution is
again seen to be the generalized eigenvector of the pencil $(\mathbf{a}_b \mathbf{a}_b^H,\frac{\sigma^2}{p_b} \mathbf{I} +
\mathbf{B}_b^H \mathbf{B}_b)$ corresponding
to the largest eigenvalue. Since it is a rank-one Hermitian matrix,
the result can be explicitly presented as $\mathbf{t}_b^\prime = \mathbf{G}_b^\perp \frac{(\frac{\sigma^2}{p_b} \mathbf{I} +
\mathbf{B}_b^H \mathbf{B}_b)^{-1} \mathbf{a}_b}{||(\frac{\sigma^2}{p_b}
\mathbf{I} + \mathbf{B}_b^H \mathbf{B}_b)^{-1} \mathbf{a}_b||}$, which is the
same result as in the rate maximization problem.  Similarly, for the second phase,
we also have the same beamformer as \eqref{eq18}.
Considering the transmit power of all the nodes, we can
now formulate the optimization problem under the global transmit power
constraint as
\begin{equation} \label{eq20}
\min_{\mathbf{p}} \quad  \max (p_{a}+\tilde{p_b},p_r+\tilde{p}_{a2}) \qquad
\mathrm{s.t.} \quad g(\mathbf{p}) \le \frac{|\mathbf{h}_{ar} \mathbf{t}_{a} |^2}{ 2^{2R_0} \sigma^2}
\end{equation}
where $g(\mathbf{p})$ is given in \eqref{eq19}. This is also a GP problem. To minimize
individual transmit powers, \eqref{eq20} should be rewritten as $\min_{\mathbf{p}}
\max(p_{a},\tilde{p_b},p_r,\tilde{p}_{a2})$ instead.

\subsection{Multiple data stream relaying}
Since ECSI is known to the relay network, Alice and the Relay can utilize certain beamformers to perform multiple data-stream relay transmission and reduce information leakage to Eve as well. The GSVD has been employed for the  traditional MIMO wiretap channel \cite{Khisti_Gaussian07}, and it operates by dividing the channels from the transmitter to the intended receiver and the eavesdropper into a set of parallel subchannels.
\begin{definition}[GSVD Transform]
Given two matrices $\mathbf{H}_{1} \in \mathbb{C}^{N_r \times N_a}$ and $\mathbf{H}_{2} \in \mathbb{C}^{N_e \times N_a}$ and $k=rank\{[\mathbf{H}_{1}^H, \mathbf{H}_{2}^H]^H\}$, there exist unitary matrices $\mathbf{U} \in \mathbb{C}^{N_r \times N_r}$, $\mathbf{V} \in \mathbb{C}^{N_e \times N_e}$ and $\mathbf{\Psi} \in \mathbb{C}^{N_a \times N_a}$, and a non-singular upper-triangular matrix $\mathbf{R} \in \mathbb{C}^{k \times k}$ such that
\begin{equation}
\mathbf{U}^H \mathbf{H}_{1} \mathbf{\Psi} = \mathbf{S}_{1} [\mathbf{R}, \mathbf{0}_{k\times N_a - k} ],\quad \mathbf{V}^H \mathbf{H}_{2} \mathbf{\Psi}= \mathbf{S}_{2} [\mathbf{R}, \mathbf{0}_{k\times N_a - k}]  \notag
\end{equation}
where $\mathbf{S}_{1} \in \mathbb{R}^{N_r \times k}$, $\mathbf{S}_{2} \in \mathbb{R}^{N_e \times k}$ are nonnegative diagonal matrices with $\mathbf{S}_{1}^T \mathbf{S}_{1} + \mathbf{S}_{2}^T \mathbf{S}_{2} = \mathbf{I}_k$, the diagonal elements of $(\mathbf{S}_{1}^T \mathbf{S}_{1})^{\frac{1}{2}}$ are ordered as $0 \le s_{1,1} \le \cdots \le s_{1,k}$, and the diagonal elements of $(\mathbf{S}_{2}^T \mathbf{S}_{2})^{\frac{1}{2}}$ are ordered as $s_{2,1} \ge \cdots \ge s_{2,k} \ge 0$.
\end{definition}

It has been shown \cite{Khisti_Secure10} that, for the standard
Gaussian MIMO wiretap channel, using the GSVD-based beamformer
\begin{equation}
\mathbf{T} = \frac{\mathbf{\Psi}}{||\mathbf{R}^{-1}||}
\left[
\begin{array}{c}
\mathbf{R}^{-1} \\
\mathbf{0}_{N_a - k \times k}
\end{array}
\right] \notag
\end{equation}
to transmit the desired signals along dimensions where $s_{1,i} \ge
s_{2,i}$ achieves the secrecy capacity in the high SNR regime with
uniform power allocation.  In this section, two transmission
strategies based on the GSVD will be investigated for the two-hop
relay channel.  In the first strategy, each transmission phase is
treated as a standard wiretap channel, and Alice and the Relay will
use GSVD-based transmit beamformers in the first and the second phase,
respectively, without any cooperative jamming from inactive nodes. In
the second strategy, a cooperative jamming scheme is proposed in which
Bob and Alice also transmit jamming signals based on the GSVD
transform in a reverse manner.

\subsubsection{Simple GSVD-based relaying}\label{sec:simp_GSVD}
To begin, we consider the case where GSVD-based beamforming is used without jamming. According to Definition 1, the MIMO channels in phase 1 and phase 2 can be decomposed as
\begin{equation*}
\mathbf{H}_{ar}=\mathbf{U}_a \mathbf{S}_{ar} [\mathbf{R}_a, \mathbf{0}_{s \times N_a-s}] \mathbf{\Psi}_a^H \qquad \mathbf{H}_{ae}=\mathbf{V}_a \mathbf{S}_{ae} [\mathbf{R}_a, \mathbf{0}_{s \times N_a-s}] \mathbf{\Psi}_a^H
\end{equation*}
\begin{equation*}
\mathbf{H}_{rb}=\mathbf{U}_r \mathbf{S}_{rb} [\mathbf{R}_r, \mathbf{0}_{s \times N_r-s}] \mathbf{\Psi}_r^H \qquad \mathbf{H}_{re}=\mathbf{V}_r \mathbf{S}_{re} [\mathbf{R}_r, \mathbf{0}_{s \times N_r-s}] \mathbf{\Psi}_r^H
\end{equation*}
where $s=min(rank\{[\mathbf{H}_{ar}^H, \mathbf{H}_{ae}^H]^H\},rank\{[\mathbf{H}_{rb}^H, \mathbf{H}_{re}^H]^H\})$, representing the maximum possible number of data streams. Alice and the Relay transmit signals with the following two beamformers respectively,
\begin{equation} \label{eq:tbfgsvd}
\mathbf{T}_a = \frac{\mathbf{\Psi}_a}{||\mathbf{R}_a^{-1}||}
\left[
\begin{array}{c}
\mathbf{R}_a^{-1} \\
\mathbf{0}_{N_a-s \times s}
\end{array}
\right] \qquad
\mathbf{T}_r = \frac{\mathbf{\Psi}_r}{||\mathbf{R}_r^{-1}||}
\left[
\begin{array}{c}
\mathbf{R}_r^{-1} \\
\mathbf{0}_{N_r-s \times s}
\end{array}
\right].
\end{equation}

\begin{proposition} \label{simplegsvd}
When \eqref{eq:tbfgsvd} is used for transmit beamforming, the secrecy rate under the simple GSVD-based relaying scheme can be expressed as:
\begin{equation}\label{eq:cap_gsvd}
R_{gsvd} = \frac{1}{2} \log_2 \frac{\min \left \{ \prod_{i=1}^s  \left(1+\frac{p_{a,i} s_{ar,i}^2}{\sigma^2 ||\mathbf{R}_a^{-1}||^2} \right),\prod_{i=1}^s \left(1+\frac{p_{r,i} s_{rb,i}^2}{\sigma^2 ||\mathbf{R}_r^{-1}||^2} \right)	\right \}}{\prod_{i=1}^s \left( 1 + \frac{p_{a,i} s_{ae,i}^2}{\sigma^2 ||\mathbf{R}_a^{-1}||^2}+
					     \frac{p_{r,i} s_{re,i}^2}{\sigma^2 ||\mathbf{R}_r^{-1}||^2} \right)}
\end{equation}
where $p_{a,i}$ and $p_{r,i}$ are the transmit power for the $i$th parallel channel from Alice and the Relay, respectively.
\end{proposition}

The proof of Proposition \ref{simplegsvd} is given in Appendix \ref{sec:appspgsvd}. Next, we will investigate the power allocation for the above transmission scheme. Maximizing the rate in (\ref{eq:cap_gsvd}) is generally a nonconvex optimization problem. However, applying the \emph{single condensation method} for GP \cite{Chiang_Power07}, the posynomial in the numerator of (\ref{eq:cap_gsvd}) can be accurately approximated as a monomial, and we can still solve this nonconvex problem through a \emph{series} of GPs.
\begin{lemma}
Let
\begin{equation} \label{eq:al1-1}
\prod_{i=1}^s f_i(p_{a,i})=\prod_{i=1}^s \left( 1 + \frac{p_{a,i} s_{ar,i}^2}{\sigma^2 ||\mathbf{R}_a^{-1}||^2} \right).
\end{equation}
We have
\begin{equation} \label{eq:al1-3}
\prod_{i=1}^s f_i(p_{a,i}) \ge \prod_{i=1}^s \widetilde{f}_i(p_{a,i}) = \prod_{i=1}^s \left(\frac{1}{\alpha_{1,i}}\right)^{\alpha_{1,i}} \left(\frac{p_{a,i} s_{ar,i}^2}{\alpha_{2,i} \sigma^2 ||\mathbf{R}_a^{-1}||^2}\right)^{\alpha_{2,i}}
\end{equation}
where $\alpha_{1,i},\alpha_{2,i}\ge0$. The inequality becomes an equality when $\alpha_{1,i},\alpha_{2,i}$ satisfy
\begin{equation}\label{eq:alpha_k}
\alpha_{1,i} = \cfrac{1}{f_i(p_{a,i})} \ , \quad
\alpha_{2,i} = \cfrac{p_{a,i}}{f_i(p_{a,i})} \cfrac{\partial f_i(p_{a,i})}{\partial p_{a,i} } \ ,
\end{equation}
in which case $\prod_{i=1}^s \widetilde{f}_i(p_{a,i})$ is the best local monomial approximation of $\prod_{i=1}^s f_i(p_{a,i})$ near $p_{a,i}$.
\end{lemma}

\begin{IEEEproof}
We can rewrite $\prod_{i=1}^s f_i(p_{a,i})$ as
\begin{align}
\prod_{i=1}^s \left( 1 + \frac{p_{a,i} s_{ar,i}^2}{\sigma^2 ||\mathbf{R}_a^{-1}||^2} \right) &=
\prod_{i=1}^s \left( \alpha_{1,i} \frac{1}{\alpha_{1,i}} + \alpha_{2,i} \frac{p_{a,i} s_{ar,i}^2}{\alpha_{2,i} \sigma^2 ||\mathbf{R}_a^{-1}||^2} \right) \\
&\ge \prod_{i=1}^s \left(\frac{1}{\alpha_{1,i}}\right)^{\alpha_{1,i}} \left(\frac{p_{a,i} s_{ar,i}^2}{\alpha_{2,i} \sigma^2 ||\mathbf{R}_a^{-1}||^2}\right)^{\alpha_{2,i}} \label{eq:agmineq}
\end{align}
where (\ref{eq:agmineq}) holds according to the arithmetic-geometric mean inequality. Noting that $\alpha_{1,i}$ and $\alpha_{2,i}$ are both positive coefficients and $\alpha_{1,i}+\alpha_{2,i}=1, \forall i$, the proof of equality is straightforward by inserting (\ref{eq:alpha_k})	 back into $\prod_{i=1}^s \widetilde{f}_i(p_{a,i})$.
\end{IEEEproof}

Similarly for the second phase, given the posynomial
\begin{equation} \label{eq:al1-2}
\prod_{i=1}^s g_i(p_{r,i})=
\prod_{i=1}^s \left( 1 + \frac{p_{r,i} s_{rb,i}^2}{\sigma^2 ||\mathbf{R}_r^{-1}||^2} \right),
\end{equation}
we have the approximation
\begin{equation} \label{eq:al1-4}
\prod_{i=1}^s \widetilde{g}_i(p_{r,i}) =
\prod_{i=1}^s \left(\frac{1}{\beta_{1,i}}\right)^{\beta_{1,i}} \left(\frac{p_{r,i} s_{rb,i}^2}{ \beta_{2,i} \sigma^2 ||\mathbf{R}_r^{-1}||^2}\right)^{\beta_{2,i}}
\end{equation}
where
\begin{equation}\label{eq:beta_k}
\beta_{1,i} = \cfrac{1}{g_i(p_{r,i})} \ ,
\beta_{2,i} = \cfrac{p_{r,i}}{g_i(p_{r,i})} \cfrac{\partial g_i(p_{r,i})}{\partial p_{r,i}} \ .
\end{equation}
The approach corresponding to these results is outlined in the following algorithm:

\textbf{Algorithm 1:} Single condensation method for power allocation

\textbf{Initialize} $p_{a,i}^{(0)}$ and $p_{r,i}^{(0)}$, $i=\{1,\cdots,s\}$.

\textbf{For} iteration $k$:
\begin{enumerate}
			\item Evaluate posynomial $f_i(p_{a,i}^{(k-1)})$ and $g_i(p_{r,i}^{(k-1)})$, according to (\ref{eq:al1-1}) and (\ref{eq:al1-2}).
			\item Compute $\alpha^{(k)}$ and $\beta^{(k)}$:
			\begin{equation}
			\left \{ \begin{array}{l}
			\alpha_{1,i}^{(k)} = \cfrac{1}{f_i(p_{a,i}^{(k-1)})} \ , \quad \beta_{1,i}^{(k)} = \cfrac{1}{g_i(p_{r,i}^{(k-1)})} \\
			\alpha_{2,i}^{(k)} = \cfrac{p_{a,i}^{(k-1)}}{f_i(p_{a,i}^{(k-1)})} \cfrac{\partial 	
			f_i(p_{a,i}^{(k-1)})}{\partial p_{a,i}^{(k-1)} } \ , \quad \beta_{2,i}^{(k)} = 		
			\cfrac{p_{r,i}^{(k-1)}}{g_i(p_{r,i}^{(k-1)})} \cfrac{\partial 				
			g_i(p_{r,i}^{(k-1)})}{\partial p_{r,i}^{(k-1)} } \ .
			\end{array} \right.
			\end{equation}

			\item Condense posynomials $f_i$ and $g_i$ into monomials $\widetilde{f}_i$ and $\widetilde{g}_i$, according to (\ref{eq:al1-3}) and (\ref{eq:al1-4}).
			\item Solve the GP
			\begin{subequations}
				\begin{align}
				\min_{\mathbf{p}}&  \max \quad \left \{ \prod_{i=1}^s \widetilde{f}_i(p_{a,i})^{-1},\prod_{i=1}^s \widetilde{g}_i(p_{r,i})^{-1}	 \right \} \prod_{i=1}^s \left ( 1 + \frac{p_{a,i} s_{ae,i}^2}{\sigma^2 ||\mathbf{R}_a^{-1}||^2}+ \frac{p_{r,i} s_{re,i}^2}{\sigma^2 ||\mathbf{R}_r^{-1}||^2} \right ) \label{eq:ag1ob} 	\\			
				\mathrm{s.t.}& \quad \sum_{i=1}^s p_{a,i} \le P, \quad \sum_{i=1}^s p_{r,i} \le P .
				\end{align}
				\end{subequations}
				\item Apply the resulting $p_{a,i}^{(k)}$ and $p_{r,i}^{(k)}$ to step 1 and loop until convergence.
\end{enumerate}

The GP problems in this successive optimization method can be solved using interior-point methods with polynomial-time complexity \cite{Boyd_tutorial07}, and it has been proven in \cite{Chiang_Power07} that the solution obtained using successive approximations for the single condensation method will efficiently converge to a point satisfying the KKT conditions of the original problem, and the global optimum can consequently be obtained. Note that \eqref{eq:ag1ob} is refered to as a \textit{generalized posynomial} \cite{Boyd_tutorial07} since it is formed from posynomials using a maximum operation, and can be easily converted to the standard posynomial form as
\begin{subequations}
\begin{align}
				\min_{\mathbf{p},\mu}&  \quad \mu ~ \prod_{i=1}^s \left ( 1 + \frac{p_{a,i} s_{ae,i}^2}{\sigma^2 ||\mathbf{R}_a^{-1}||^2}+ \frac{p_{r,i} s_{re,i}^2}{\sigma^2 ||\mathbf{R}_r^{-1}||^2} \right ) \label{eq:ag1ob2} 	\\			
				\mathrm{s.t.}& \quad \sum_{i=1}^s p_{a,i} \le P, \quad \sum_{i=1}^s p_{r,i} \le P \\
				             & \quad \prod_{i=1}^s \widetilde{f}_i(p_{a,i})^{-1} \mu^{-1} \le 1, \quad \prod_{i=1}^s \widetilde{g}_i(p_{r,i})^{-1} \mu^{-1} \le 1.
\end{align}
\end{subequations}

\subsubsection{GSVD-based PCJ}\label{sec:GSVDPCJ}
A GSVD-based, partial cooperative jamming scheme is proposed in this subsection. In this case, Alice and the Relay will still use the same transmit beamformers as in the case without cooperative jamming. Since Bob and Alice are normally inactive in phase 1 and phase 2 respectively, they can act as temporary helpers to help improve the secrecy rate.  As before, however, the power used for jamming must come from the total power budget of $P$ in each hop.  A GSVD-based beamformer for the jamming signal is used by Bob in the first phase, due to the assumption that ECSI is available. The GSVD is implemented in a reverse fashion, since Bob in phase 1 considers Eve as the intended receiver of the jamming and wants to avoid leaking interference signals to the Relay. Similarly in phase 2, Alice treats Eve as the intended receiver. The signal model for this scheme is given in Section \ref{sec:cjmod}.

Performing the GSVD for the channels from Bob to Eve and the Relay according to Definition 1, we have
\begin{equation*}
\mathbf{H}_{be}=\mathbf{U}_b \mathbf{S}_{be} [\mathbf{R}_b,\mathbf{0}_{k_b \times N_r-k_b}] \mathbf{\Psi}_b^H \qquad \mathbf{H}_{br}=\mathbf{V}_b \mathbf{S}_{br} [\mathbf{R}_b,\mathbf{0}_{k_b \times N_r-k_b}] \mathbf{\Psi}_b^H
\end{equation*}
\begin{equation*}
\mathbf{H}_{ae}=\mathbf{U}_{a^\prime} \mathbf{S}_{ae} [\mathbf{R}_{a^\prime},\mathbf{0}_{k_a \times N_a-k_a}] \mathbf{\Psi}_{a^\prime}^H \qquad \mathbf{H}_{ab}=\mathbf{V}_{a^\prime} \mathbf{S}_{ab} [\mathbf{R}_{a^\prime},\mathbf{0}_{k_a \times N_a-k_a}] \mathbf{\Psi}_{a^\prime}^H
\end{equation*}
where $k_b=rank\{[\mathbf{H}_{be}^H, \mathbf{H}_{br}^H]^H\}$. Bob and Alice use the following jamming beamformers to implement the reverse GSVD
\begin{equation*}
\mathbf{T}_b^\prime = \frac{\mathbf{\Psi}_b}{||\mathbf{R}_b^{-1}||}
\left[
\begin{array}{c}
\mathbf{R}_b^{-1} \\
\mathbf{0}_{N_r-k_b \times k_b}
\end{array}
\right] \qquad
\mathbf{T}_a^\prime = \frac{\mathbf{\Psi}_b}{||\mathbf{R}_{a^\prime}^{-1}||}
\left[
\begin{array}{c}
\mathbf{R}_{a^\prime}^{-1} \\
\mathbf{0}_{N_a-k_a \times k_a}
\end{array}
\right],
\end{equation*}
and, unlike the simple GSVD-based relaying
scheme, there will be jamming energy present in the signals received by the Relay and Bob.
For Eve, the received signal is given by
\begin{equation}
\mathbf{y}_e=
\left[ \begin{array}{l}
\mathbf{H}_{ae} \mathbf{T}_{a} \mathbf{D}_{a}  \\
 \mathbf{H}_{re} \mathbf{T}_{r} \mathbf{D}_{r}
\end{array} \right]
\mathbf{z}
+
\left[ \begin{array}{l}
\mathbf{H}_{be} \mathbf{T}_b^\prime \mathbf{z}_b^\prime + \mathbf{n}_{e1}  \notag \\
\mathbf{H}_{ae} \mathbf{T}_a^\prime \mathbf{z}_a^\prime + \mathbf{n}_{e2}
\end{array} \right]
=\widetilde{\mathbf{H}}_{e} \mathbf{z} + \widetilde{\mathbf{n}}_e.
\end{equation}
Employing the above jamming beamformers, the mutual information between Alice and Bob is
\begin{equation}
I_d=\min \left\{\frac{1}{2} \log_2 \det (\mathbf{I}+ \mathbf{H}_{ar} \mathbf{T}_a \mathbf{Q}_{za} \mathbf{T}_a^H \mathbf{H}_{ar}^H \mathbf{Q}_{\tilde{n}r}^{-1}),\frac{1}{2} \log_2 \det (\mathbf{I}+ \mathbf{H}_{rb} \mathbf{T}_r \mathbf{Q}_{zr} \mathbf{T}_r^H  \mathbf{H}_{rb}^H \mathbf{Q}_{\tilde{n}b}^{-1})\right\}
\end{equation}
where $\mathbb{E}(\mathbf{z}_a^\prime \mathbf{z}_a^{\prime H})=\mathbf{Q}_{za^\prime}$ $\mathbb{E}(\mathbf{z}_b^\prime \mathbf{z}_b^{\prime H})=\mathbf{Q}_{zb^\prime}$, and
\begin{equation*}
\mathbf{Q}_{\tilde{n}r} = \mathbb{E} [(\mathbf{H}_{br} \mathbf{T}_b^\prime \mathbf{z}_b^\prime + \mathbf{n}_r)(\mathbf{H}_{br} \mathbf{T}_b^\prime \mathbf{z}_b^\prime + \mathbf{n}_r)^H]
											= \mathbf{H}_{br} \mathbf{T}_b^\prime \mathbf{Q}_{zb^\prime} \mathbf{T}_b^{\prime H} \mathbf{H}_{br}^H + \sigma^2 \mathbf{I}
\end{equation*}
\begin{equation*}
\mathbf{Q}_{\tilde{n}b} = \mathbb{E} [(\mathbf{H}_{ab} \mathbf{T}_a^\prime \mathbf{z}_a^\prime + \mathbf{n}_b)(\mathbf{H}_{ab} \mathbf{T}_a^\prime \mathbf{z}_a^\prime + \mathbf{n}_b)^H]
											= \mathbf{H}_{ab} \mathbf{T}_a^\prime \mathbf{Q}_{za^\prime} \mathbf{T}_a^{\prime H} \mathbf{H}_{ab}^H + \sigma^2 \mathbf{I}.
\end{equation*}
The mutual information at Eve is
\begin{align}
I_e &= \min \left\{\frac{1}{2} \log_2 \det (\mathbf{I}+ \mathbf{H}_{ar} \mathbf{T}_a \mathbf{Q}_{za} \mathbf{T}_a^H \mathbf{H}_{ar}^H \mathbf{Q}_{\tilde{n}r}^{-1}),\frac{1}{2} \log_2 \det (\mathbf{I}+\widetilde{\mathbf{H}}_e \mathbf{Q}_{z} \widetilde{\mathbf{H}}_e^H \mathbf{Q}_{\tilde{n}e}^{-1}) \right\}
\end{align}
where
\begin{equation*}
\mathbf{Q}_{\tilde{n}e} =
\left[ \begin{array}{cc}
\mathbf{H}_{be} \mathbf{T}_b^\prime \mathbf{Q}_{zb^\prime} \mathbf{T}_b^{\prime H} \mathbf{H}_{be}^H +\sigma^2 \mathbf{I} & \mathbf{0}\\
\mathbf{0} & \mathbf{H}_{ae} \mathbf{T}_a^\prime \mathbf{Q}_{za^\prime} \mathbf{T}_a^{\prime H} \mathbf{H}_{ae}^H +\sigma^2 \mathbf{I} \\
\end{array} \right] .
\end{equation*}
To maximize the secrecy rate, we then have the following optimization problem:
\begin{align}
\max_{\mathbf{Q}_{za},\mathbf{Q}_{zb^\prime},\mathbf{Q}_{zr},\mathbf{Q}_{za^\prime}}&  \quad R_{gsvd}^{PCJ} \notag \\
\mathrm{s.t.}  \quad \textrm{tr}(\mathbf{Q}_{za} + \mathbf{Q}_{zb^\prime}) \le &P,
							 \quad  \textrm{tr}(\mathbf{Q}_{zr} + \mathbf{Q}_{za^\prime}) \le P \label{eq:glo2}
\end{align}
where $R_{gsvd}^{PCJ} = I_d - I_e$ and $P$ is the global power constraint.
\begin{remark}
The secrecy rate in this case does not have a form similar to (\ref{eq:cap_gsvd}), and finding the optimal power allocation for this case is generally intractable. Therefore, we will use Newton's method initialized with the optimal point from the GSVD-based relaying algorithm. Though this may not find the global optimum, we can at least gain insight into this strategy. The global power constraints in (\ref{eq:glo2}) are set for fair comparison with the case without cooperative jamming.
\end{remark}


\section{Secure relaying with unknown ECSI} \label{sec:fixrate}
In this section, we assume that ECSI is unknown to the relay
network. Thus, Alice and the Relay can no longer use beamforming
methods like those based on the GSVD to selectively transmit
information away from and jamming signals towards the
eavesdropper. However, cooperative jamming can still be used to
improve the secrecy of the information in the two-hop network.  As
described below, the approach we take to achieve this goal is to first
meet a fixed target rate for the relay link, and then allocate all
remaining resources to wide-area jamming, while guaranteeing that the
jamming signal has no impact on the desired information.

We propose a cooperative jamming strategy in which the signal space is
divided into two orthogonal subspaces, an information subspace and
a jamming subspace. Both PCJ and FCJ approaches can be applied in this
scenario. For PCJ, any available jamming power will only be allocated
to information transmitters, while Bob (phase 1) and Alice (phase 2)
remain inactive. For FCJ, both the transmitter and the temporary
helpers can perform cooperative jamming in the jamming subspace, which
will allow the legitimate receivers to use beamforming to reject
interference from this subspace. Note that when using FCJ, cooperative
jamming requires the receiver to broadcast the jamming subspace so
that the interference can be aligned at the desired receiver without a
loss of information. Although Eve may also be aware of this subspace,
she can not remove the jamming signal since she sees different
channels from the transmitters and jammers.

In phase 1, assume  $span\{\mathbf{H}_{ar}\}=span\{\eta_1,\eta_2,\dots,\eta_k,\eta_{k+1},\dots,\eta_m\}$, where $k$ is no greater than the maximum possible number of data streams, and $\eta_1,\eta_2\,\dots,\eta_m$ form an orthonormal basis. The information and jamming subspaces are defined to be $\mathcal{S}_1$ and $\mathcal{J}_1$, respectively, where $\mathcal{S}_1=span\{\eta_1,\eta_2,\dots,\eta_k\}$ and $\mathcal{J}_1= \mathcal{S}_1^\perp$. Assuming the receive beamformer matrix at the Relay is $\mathbf{W}_r=[\eta_1,\eta_2,\dots,\eta_k]$, the signal received by the Relay is
\begin{equation}
\widetilde{\mathbf{y}}_r=\mathbf{W}_r^H [\mathbf{H}_{ar} (\mathbf{T}_{a} \mathbf{z}_{a} + \mathbf{T}_{a}^\prime \mathbf{z}_{a}^\prime) + \mathbf{H}_{br} \mathbf{T}_{b}^\prime  \mathbf{z}_{b}^\prime + \mathbf{n}_r]
						=\widetilde{\mathbf{H}}_{ar} \mathbf{z}_{a} + \widetilde{\mathbf{n}}_r
\end{equation}
where $\widetilde{\mathbf{H}}_{ar}=\mathbf{W}_r^H \mathbf{H}_{ar} \mathbf{T}_{a}$, $\mathbf{z}_{a}$ is the information signal vector transmitted by Alice with covariance $\mathbf{Q}_{za}$, $\mathbf{z}_{a}^\prime$ and $\mathbf{z}_{b}^\prime$ are jamming signals transmitted by Alice and Bob, with covariance matrices $\mathbf{Q}_{z^\prime a}$ and $\mathbf{Q}_{z^\prime b}$, respectively. The transmit beamformers are chosen such that $\mathbf{H}_{ar} \mathbf{T}_{a} \mathbf{z}_{a} \in \mathcal{S}_1$, and $\mathbf{H}_{ar} \mathbf{T}_{a}^\prime \mathbf{z}_{a}^\prime \in \mathcal{J}_1, \mathbf{H}_{br} \mathbf{T}_{b}^\prime  \mathbf{z}_{b}^\prime \in \mathcal{J}_1$. The signal received by Eve in phase 1 is
\begin{equation}
\mathbf{y}_{e1}=\mathbf{H}_{ae} (\mathbf{T}_{a} \mathbf{z}_{a} + \mathbf{T}_{a}^\prime \mathbf{z}_{a}^\prime) + \mathbf{H}_{be} \mathbf{T}_{b}^\prime  \mathbf{z}_{b}^\prime + \mathbf{n}_{e1}
						=\widetilde{\mathbf{H}}_{ae} \mathbf{z}_{a} + \widetilde{\mathbf{n}}_{e1}
\end{equation}
where
$
\widetilde{\mathbf{n}}_{e1} = \mathbf{H}_{ae} \mathbf{T}_{a}^\prime \mathbf{z}_{a}^\prime + \mathbf{H}_{be} \mathbf{T}_{b}^\prime  \mathbf{z}_{b}^\prime + \mathbf{n}_{e1}
$.

In phase 2, signal $\mathcal{S}_2$ and jamming $\mathcal{J}_2$ subspaces are chosen from $span\{\mathbf{H}_{rb}\}$, and similar to phase 1, the signals at Bob and Eve are
\begin{align}
\widetilde{\mathbf{y}}_b&=\mathbf{W}_b^H [\mathbf{H}_{rb} (\mathbf{T}_{r} \mathbf{z}_{r} + \mathbf{T}_{r}^\prime \mathbf{z}_{r}^\prime) + \mathbf{H}_{ab} \mathbf{T}_{a2}^\prime  \mathbf{z}_{a2}^\prime + \mathbf{n}_b]
						=\widetilde{\mathbf{H}}_{rb} \mathbf{z}_r + \widetilde{\mathbf{n}}_b\\
\mathbf{y}_{e2}&=\mathbf{H}_{re} (\mathbf{T}_{r} \mathbf{z}_{r} + \mathbf{T}_{r}^\prime \mathbf{z}_{r}^\prime) + \mathbf{H}_{ae} \mathbf{T}_{a2}^\prime  \mathbf{z}_{a2}^\prime + \mathbf{n}_{e2}
						=\widetilde{\mathbf{H}}_{re} \mathbf{z}_r + \widetilde{\mathbf{n}}_{e2}
\end{align}
where $\widetilde{\mathbf{H}}_{rb}=\mathbf{W}_b^H \mathbf{H}_{rb} \mathbf{T}_{r}$, $\widetilde{\mathbf{H}}_{re}=\mathbf{H}_{re} \mathbf{T}_{r}$ and
$\widetilde{\mathbf{n}}_{e2} = \mathbf{H}_{re} \mathbf{T}_{r}^\prime \mathbf{z}_{r}^\prime + \mathbf{H}_{ae} \mathbf{T}_{a2}^\prime  \mathbf{z}_{a2}^\prime + \mathbf{n}_{e2}$, $\mathbf{z}_r$ is the information signal transmitted by the Relay with covariance $\mathbf{Q}_{zr}$, $\mathbf{z}_{a2}^\prime$ and $\mathbf{z}_r^\prime $ are jamming signals transmitted by Alice and the Relay, with covariance matrices $\mathbf{Q}_{z^\prime a2}$ and $\mathbf{Q}_{z^\prime r}$, respectively. As before, the beamformers $\mathbf{T}_r$ and $\mathbf{T}_r^\prime$ force $\mathbf{H}_{rb} \mathbf{T}_{r} \mathbf{z}_{r} \in \mathcal{S}_2$, and $\mathbf{H}_{rb} \mathbf{T}_{r}^\prime  \mathbf{z}_{r}^\prime \in \mathcal{J}_2$, $\mathbf{H}_{ab} \mathbf{T}_{a2}^\prime  \mathbf{z}_{a2}^\prime \in \mathcal{J}_2$.

The cooperative scheme outlined in this section involves the allocation of power and the number of dimensions for the information and jamming subspaces. If the MIMO channel is rich enough, more dimensions allocated to the signal subspace increases the amount of power available for jamming, but leads to a smaller dimensional jamming subspace for both transmitters and cooperative jammers. More antennas for Eve usually requires a higher dimensional jamming subspace, especially when ECSI is unknown to the transmitters. One of the advantages of FCJ in this case is that in addition to the pre-assigned jamming subspace of dimension $N_a-k$ (for phase 1), the helpers provide jamming support in additional dimensions due to the fact they have different channels to Eve. Taking the transmission in phase 1 as an example, assuming $k$ dimensions are assigned to the information subspace, the jamming subspace seen from Eve will be greater than $N_a-k$.  In particular, $N_a-k \le dim(span\{\mathbf{H}_{ae} \mathbf{T}_{a}^\prime \} \cap span\{ \mathbf{H}_{be} \mathbf{T}_{b}^\prime \}) \le 2(N_a-k)$.

Therefore, the tradeoff between power and allocation of the jamming subspace dimension needs to be considered. In this case, we propose to use an approach similar to that in \cite{Mukherjee_Fixed-rate09} and minimize the product of the power allocated to the information signal and the dimension of the information subspace, $(p_{a}+p_{r})k$, such that the fixed target rate for the relay transmission is achieved. We then allocate all the remaining dimensions and power for jamming. Since the ECSI is not known, the jamming power will be uniformly distributed among all available dimensions at the transmitters and cooperative jammers. Assuming the target rate for the relay transmission is $R_t$, we have the following FCJ algorithm:

\textbf{Algorithm 2:} FCJ with unknown ECSI
\begin{enumerate}
\item \textbf{Initialize} $svd(\mathbf{H}_{ar})=\mathbf{U}_{ar} \mathbf{\Sigma}_{ar} \mathbf{V}_{ar}^H$ and $svd(\mathbf{H}_{rb})=\mathbf{U}_{rb} \mathbf{\Sigma}_{rb} \mathbf{V}_{rb}^H$.
\item \textbf{While} $i \le s$
	\begin{itemize}
		\item Let $\mathbf{W}_r= \mathbf{U}_{ar}[:,1:i]$, $\mathbf{W}_b=\mathbf{U}_{rb}[:,1:i]$, $\mathbf{T}_{a}=\mathbf{V}_{ar}[:,1:i]$, $\mathbf{T}_r=\mathbf{V}_{rb}[:,1:i]$.
		\item Let $\mathbf{T}_a^\prime= \mathbf{V}_{ar}[:,i+1:N_a]$, $\mathbf{T}_r^\prime=\mathbf{V}_{rb}[:,i+1:N_r]$.
		\item Let $svd(\mathbf{W}_r^H \mathbf{H}_{br})=\mathbf{U}_{br} \mathbf{\Sigma}_{br} \mathbf{V}_{br}^H$, $\mathbf{T}_{b}^\prime=\mathbf{V}_{br}[:,i+1:N_b]$, and $svd(\mathbf{W}_b^H \mathbf{H}_{ab})=\mathbf{U}_{ab} \mathbf{\Sigma}_{ab} \mathbf{V}_{ab}^H$, $\mathbf{T}_{a2}^\prime=\mathbf{V}_{ab}[:,i+1:N_a]$.
	\item Solve the following problem
				\begin{align*}
					&p_{a}^{(i)}=\min \  \textrm{tr}(\mathbf{Q}_{za}), \quad p_{r}^{(i)}=\min \ \textrm{tr}(\mathbf{Q}_{zr}) \\
					&\mathrm{s.t.} \ \frac{1}{2} \log_2 \det (\mathbf{I}+ \frac{1}{\sigma^2} \widetilde{\mathbf{H}}_{ar} \mathbf{Q}_{za} \widetilde{\mathbf{H}}_{ar}^H) = R_t, \quad \frac{1}{2} \log_2 \det (\mathbf{I}+\frac{1}{\sigma^2} \widetilde{\mathbf{H}}_{rb} \mathbf{Q}_{zr} \widetilde{\mathbf{H}}_{rb}^H) = R_t
				\end{align*}
				where the water filling algorithm is used to determine $\mathbf{Q}_{za}$ and $\mathbf{Q}_{zr}$.
\end{itemize}
\item Find $k= \arg \min_i ~ [ p_{a}^{(i)}+p_{r}^{(i)} ] \cdot i$, and determine all beamformers for the resulting $k$.
\item Allocate $p_{a}^{(k)}$ to $diag\{\mathbf{Q}_{za}\}$, and $p_{r}^{(k)}$ to $diag\{\mathbf{Q}_{zr}\}$ using water filling.
\item Uniformly allocate $P-p_{a}^{(k)}$ to $diag\{\mathbf{Q}_{z^\prime a},\mathbf{Q}_{z^\prime b}\}$, and $P-p_{r}^{(k)}$ to $diag\{\mathbf{Q}_{z^\prime r},\mathbf{Q}_{z^\prime a2}\}$.
\end{enumerate}

The PCJ algorithm in the unknown-ECSI case is similar to that for FCJ, except that jamming support will not be provided by Bob (in phase 1) and Alice (in phase 2), and thus the beamformers $\mathbf{T}_b^\prime$ and $\mathbf{T}_{a2}^\prime$ in step 2 will not be used. In step 5, when the necessary amount of power for information signals is assigned, all remaining jamming power will be used by Bob and Alice in phase 1 and phase 2, respectively; \textit{i.e.}, the power $P-p_{a}^{(k)}$ and $P-p_{r}^{(k)}$ will instead be assigned to $diag\{\mathbf{Q}_{z^\prime a}\}$ and $diag\{\mathbf{Q}_{z^\prime r}\}$. In either approach, the optimization problem in step 2 can be solved with a simple line search. If the minimum rate $R_t$ can not be achieved with the available power, the link is assumed to be in outage. In this algorithm, we assume that the Relay uses the same information dimension as Alice, as discussed in Section \ref{sec:sm}. However, using different information dimensions for the two phases with a more complicated coding scheme may also be an interesting case to consider for future work. 


\section{Numerical Results} \label{sec:nr} 
\begin{figure}[ht]
\centering
\includegraphics[width=0.55\textwidth]{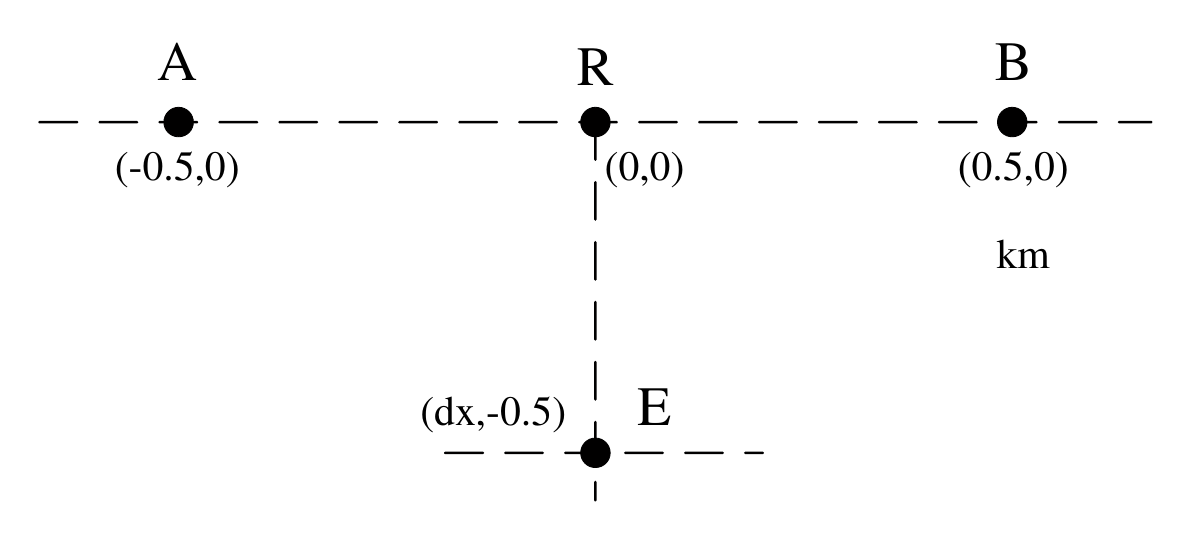}
\caption{\label{fig:sim_mod} Simulation scenario showing locations for Alice(A), Bob(B), Relay(R) and Eve(E).}
\end{figure}

In the following
simulations, the elements of all the channel matrices are assumed to
be i.i.d. complex Gaussian. As shown in Fig.~\ref{fig:sim_mod}, Alice,
Bob, the Relay and Eve are assumed to be located at $(-0.5,0)$,
$(0.5,0)$, $(0,0)$, $(d_x,-0.5)$ respectively, where distances are
expressed in kilometers. We adopt a simple transmission model in which
the standard deviation of each channel coefficient is inversely
proportional to the distance between two nodes.  We assume a path-loss
coefficient of $3$, and the same background noise power
$\sigma^2=-60$dBm at all nodes.  All results are calculated based on
an average of 3000 independent trials.

For the known-ECSI case discussed in Section \ref{sec:gsvd}, we
examine the performance of the following three schemes: PCJ for single
data stream relaying (Section~\ref{sec:single_data}), simple
GSVD-relaying (Section~\ref{sec:simp_GSVD}) and also GSVD-PCJ
(Section~\ref{sec:GSVDPCJ}) for multiple data streams. For the unknown
ECSI case discussed in Section \ref{sec:fixrate}, both the FCJ and PCJ
approaches are simulated.  For each of these schemes, we also examine
the impact of both global and individual power constraints.  For the
case of individual power constraints, we assume the total transmit
power to be evenly distributed to all transmit nodes. Furthermore, in
order to examine the performance gain of the proposed cooperative
jamming schemes and optimization algorithms, we also investigate cases
using uniform power allocation, as well as cases involving conventional relay
transmissions without jamming.

\begin{figure}[ht]
\centering
\includegraphics[width=0.6\textwidth]{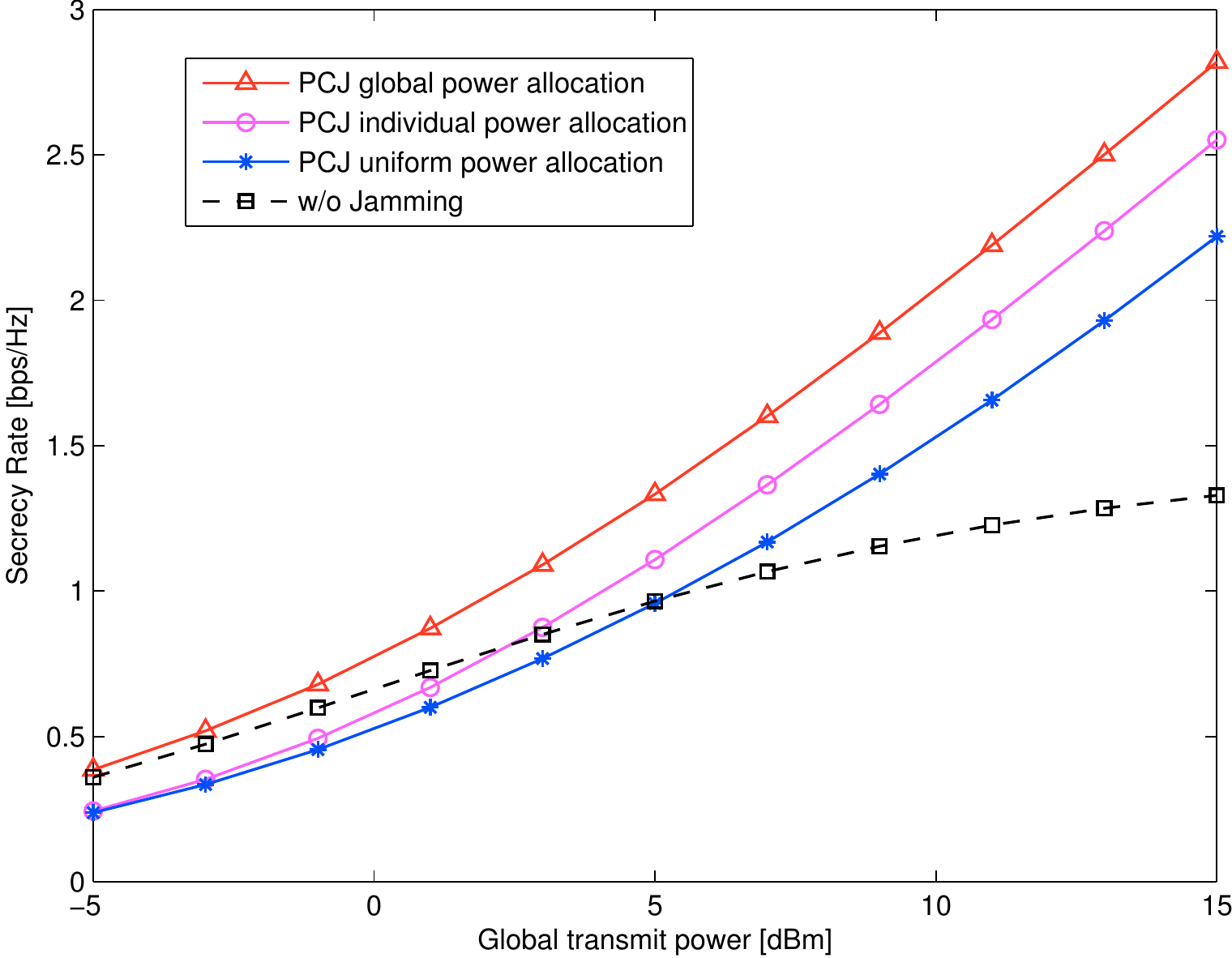}
\caption{\label{fig:mene1} Secrecy rate vs. transmit power for $N_a=4, N_b=4, N_e=1, N_r=1$, and Eve located at $(0,-0.5)$.}
\end{figure}

The secrecy rate as a function of transmit power is shown in
Fig.~\ref{fig:mene1} for a case with known ECSI, where Alice and Bob
both have four antennas, and the Relay and Eve each has one. Eve is
assumed to be located closer to the Relay at $(0,-0.5)$, which (as
will be seen in the next example) is usually the worst-case assumption
for the relay link.  This will be the default assumption unless
otherwise specified.  Compared to traditional DF relaying, the PCJ
schemes provide a significant improvement in terms of secrecy rate in
the medium and high SNR regime.  The benefit of having the flexibility
associated with a global power constraint over fixed individual power
constraints is clearly evident. Also, the performance gain of using
geometric programming for power allocation is obvious, compared to the
uniform power allocation scheme. We can also see that even the conventional
relaying scheme is better than PCJ schemes with individual or
uniform power allocation when the transmit power is low. This is because, with a less flexible power adjustment, a fraction of power that could
have brought higher secrecy rate if used for data transmission is
wasted on jamming signals. This illustrates the importance of an efficient
power allocation if cooperative jamming support is applied.

\begin{figure}[ht]
\centering
\includegraphics[width=0.6\textwidth]{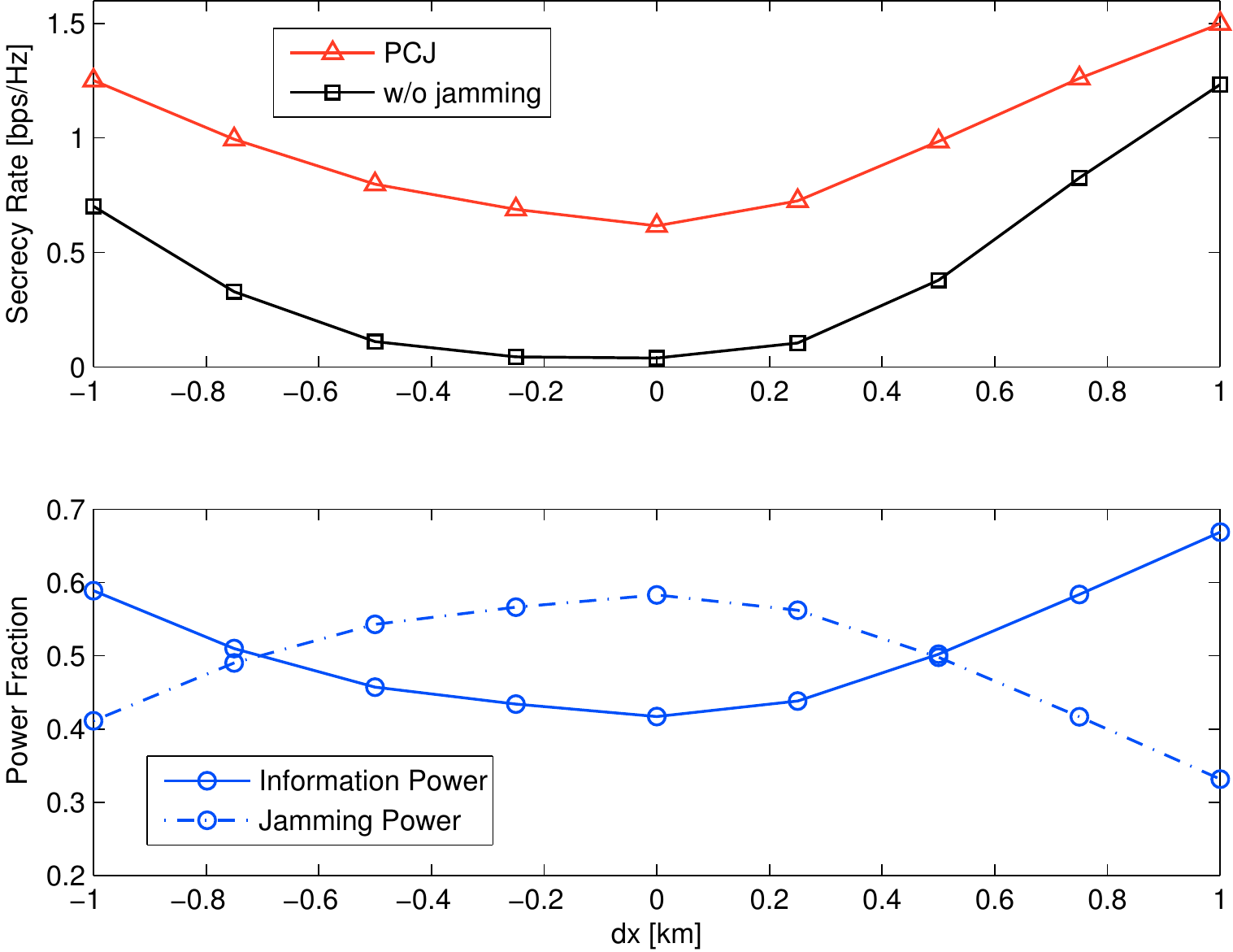}
\caption{\label{fig:mefrac} Secrecy rate and transmit power fraction
vs. eavesdropper location for $N_a=4, N_b=4, N_e=4, N_r=1$, global power constraint $P=10$dBm, and Eve's location
varies from $(-1,-0.5)$ to $(1,-0.5)$.}
\end{figure}

Fig.~\ref{fig:mefrac} presents the impact of Eve's location on the
transmit power fraction for both the information and jamming signals,
assuming that the global transmit power is limited to $10$dBm. Unlike
the settings in Fig.~\ref{fig:mene1}, Eve has four antennas in this
scenario, which provides her with increased eavesdropping
abilities. In this case, we plot the secrecy rate performance as Eve
moves from $(-1,-0.5)$ to $(1,-0.5)$.  The secrecy rate is smallest
when Eve is at the midpoint $(0,-0.5)$, and increases in either
direction away from $(0,-0.5)$.  Note also that the fraction of the
transmit power devoted to jamming also decreases as Eve moves away
from the midpoint.  This behavior is due to the fact that, when Eve is
closer to either Alice or Bob, most of her information about the
desired signal comes from only one of the hops, due to the fact that
the other hop is farther away and can be effectively jammed with
minimal power by the transmitter she is closest to.  This is the
primary benefit of the cooperative jamming support provided by PCJ.

\begin{figure}[ht]
\centering
\includegraphics[width=0.6\textwidth]{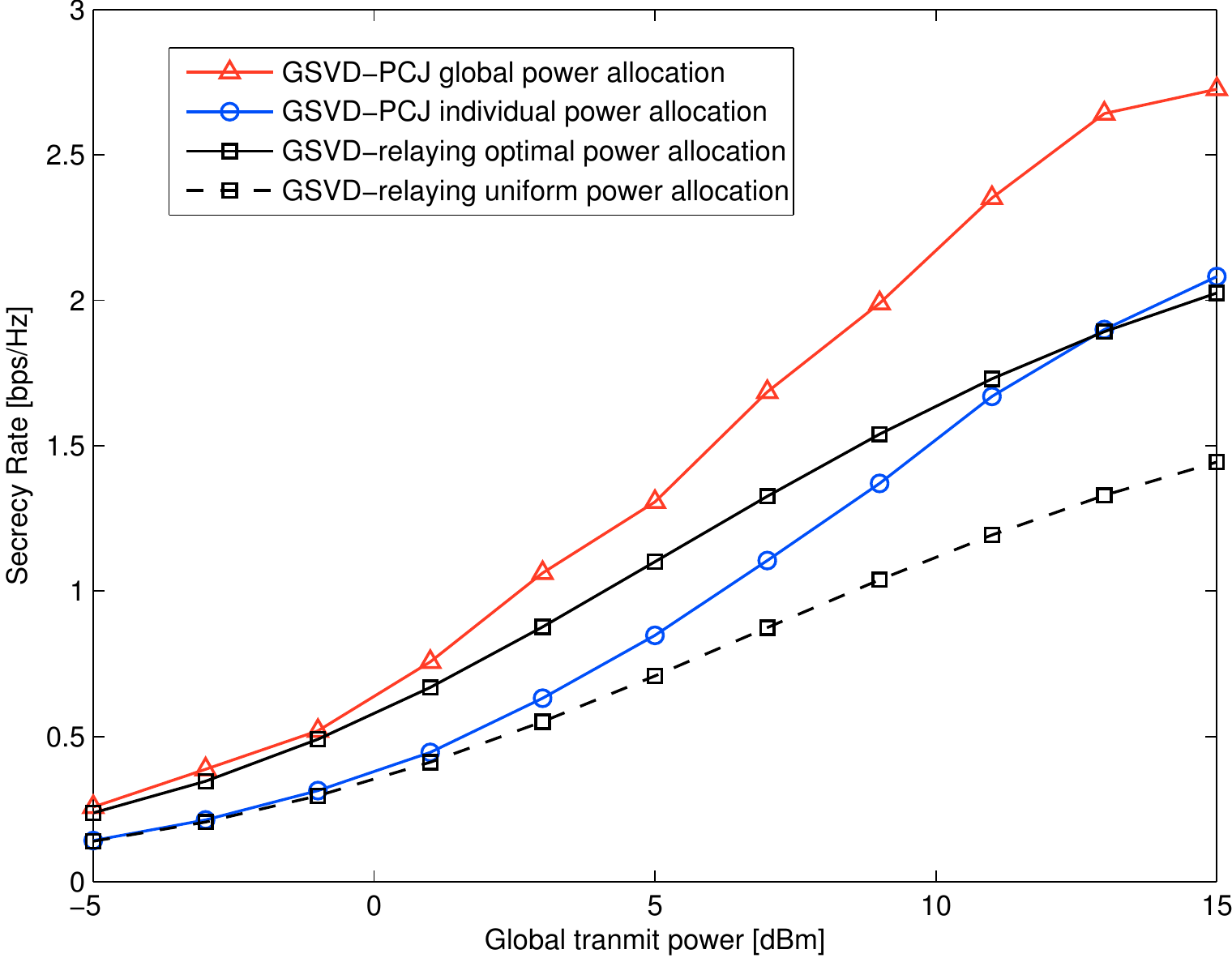}
\caption{\label{fig:mrne4} Secrecy rate vs. transmit power for $N_a=4, N_b=4, N_e=4, N_r=4$, and
Eve located at $(0,-0.5)$.}
\end{figure}

The performance of GSVD-based relaying without cooperative jamming and
GSVD-based PCJ strategies, where the relay link transmits multiple
data streams, is shown in Fig.~\ref{fig:mrne4}.  Here we see that
cooperative jamming with global power allocation provides considerable
gain in secrecy rate over other schemes. However, the use of individual power
constraints significantly degrades the benefit of the jamming signals,
although it still approaches and even surpasses the performance of
GSVD-relaying with optimal power allocation when the transmit power is
higher. In addition, we also see the 
benefit of Algorithm 1 for power allocation in the GSVD-relaying scheme, compared with using
simple uniform power allocations.

\begin{figure}[ht]
\centering
\includegraphics[width=0.6\textwidth]{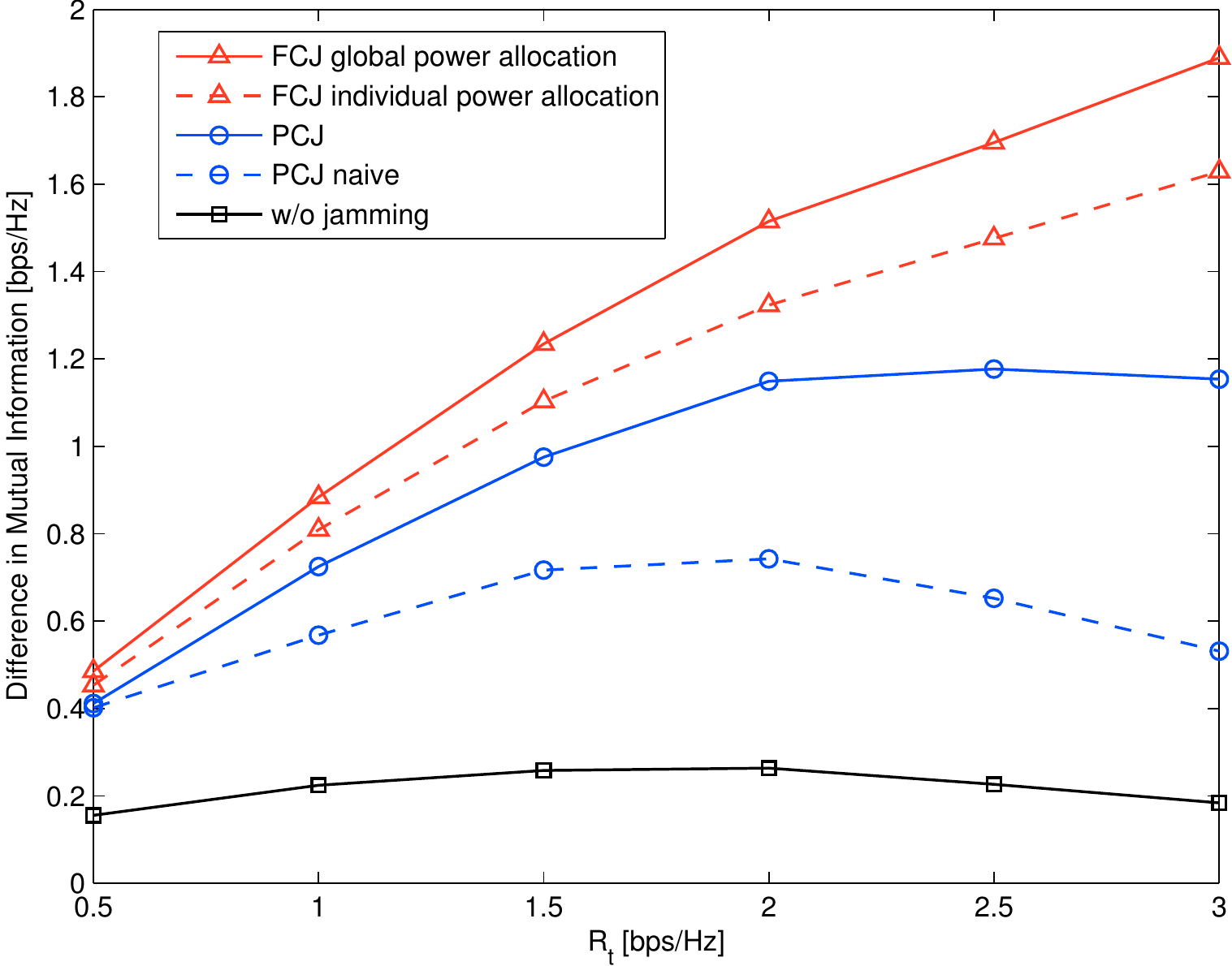}
\caption{\label{fig:mranne4} Secrecy performance vs. rate constraint for relay link when ECSI is unknown,
$N_a=4, N_b=4, N_e=4, N_r=4$, Eve located at $(0,-0.5)$, $P=15$dBm.}
\end{figure}

Finally we consider examples for the case where ECSI is not
available. The transmit power $P$ is set to be $15$dBm in these examples.  In
Fig.~\ref{fig:mranne4}, all nodes are equipped with four antennas, and
the secrecy performance is given as a function of the rate constraint at
Bob. For purpose of comparison, a naive PCJ scheme that uses the
criterion  $\min(p_a+p_r)$ (instead of $\min(p_a+p_r)k$ as discussed
in Section \ref{sec:fixrate}) is also simulated.  It can be seen that
if no jamming signals are used, there is little difference between the
mutual information at Bob and that at Eve, and thus we expect the
secrecy of the message to be low.  Similar to Fig.~\ref{fig:mene1},
the individual power constraint will reduce the secrecy performance
due to the inefficiency of the power assignment. We can see that FCJ
achieves a big performance gain compared with PCJ as $R_t$ increases,
since FCJ leads to a higher dimensional jamming subspace than PCJ,
although they transmit with the same jamming power. In addition, the performance
of PCJ begins to level off and even drop for high $R_t$,
since more power is allocated to information signals, and the
protection from eavesdropping is reduced.

\begin{figure}[ht]
\centering
\includegraphics[width=0.6\textwidth]{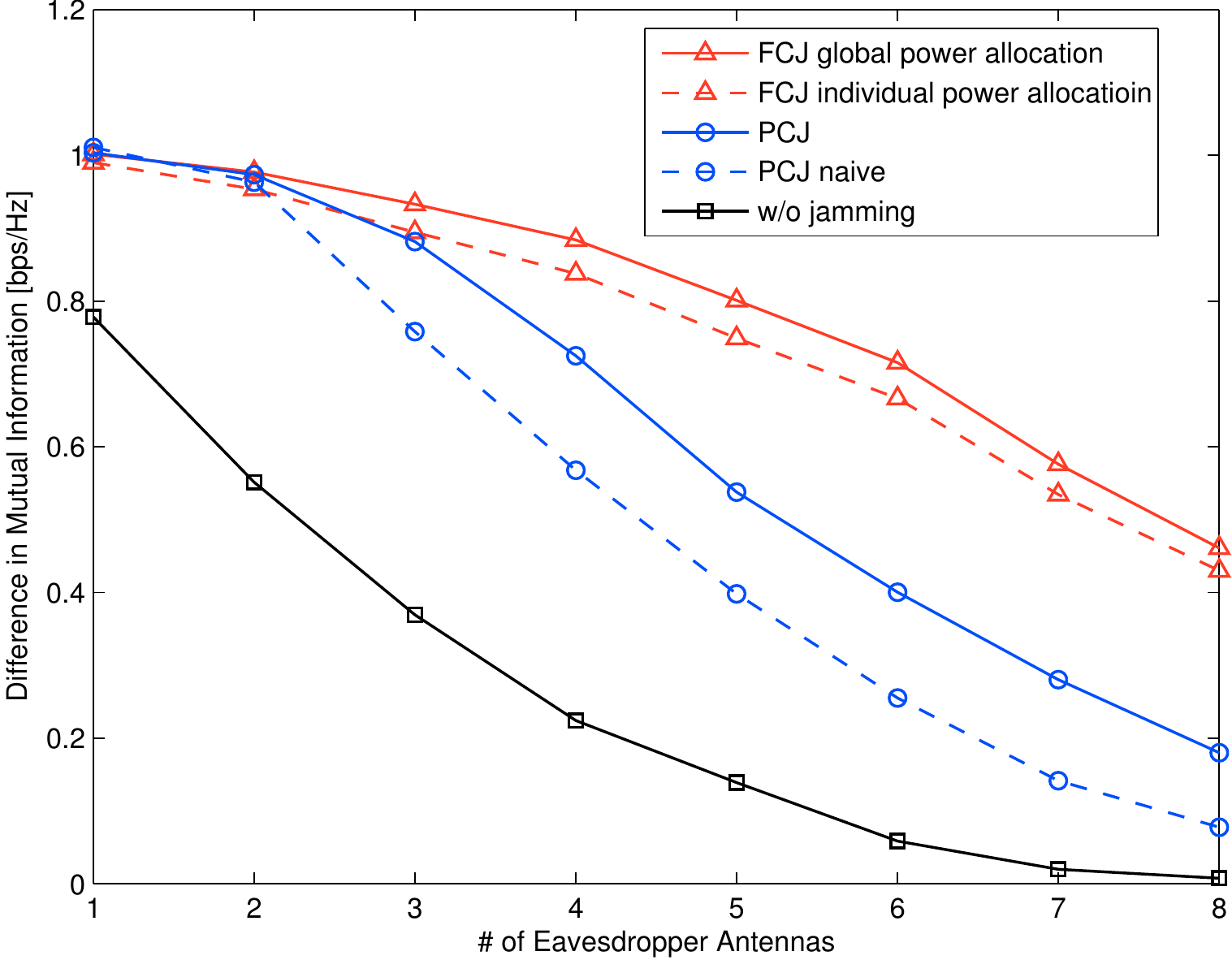}
\caption{\label{fig:mrnane18} Secrecy performance vs. eavesdropper antenna number when ECSI is unknown,
$N_a=4, N_b=4, N_r=4$, Eve located at $(0,-0.5)$, fixed target rate $R_t=1$bps/Hz, $P=15$dBm.}
\end{figure}

Fig.~\ref{fig:mrnane18} provides a detailed look at how the number of
eavesdropper antennas affects the performance of the different
cooperative jamming schemes. In this case, we fix the target rate for
relay transmission to be $R_t=1$bps/Hz. Alice, Bob and the Relay are
equipped with four antennas, and the number of Eve's antennas
increases from one to eight. It can be seen that when Eve has only one
antenna, little advantage is observed for FCJ since Eve only receives
single-dimensional signals. However, as the capability of the
eavesdropper increases (\textit{i.e.} when Eve has more antennas), the
relative gain of FCJ over PCJ increases, although the performance of
all methods decreases.

\section{Conclusions} \label{sec:con} 

In this paper, we have proposed partial cooperative jamming (PCJ) and
full cooperative jamming (FCJ) strategies for two-hop DF relay systems
in the presence of an eavesdropper that can wiretap both transmission
phases. Both single and multiple data stream transmission scenarios
were considered. For single data stream relaying, the system design
was conducted from the perspective of secrecy rate maximization and
transmit power minimization. By adopting the zero-forcing constraint
that the jamming signals be nulled out at the intended receivers, we
obtained closed-form expressions for the jamming beamformers and the
corresponding power allocation. For the case of multiple data stream
transmission, we proposed a GSVD-based relaying scheme without
jamming, as well as a GSVD-based PCJ scheme. The latter shows a
significant performance improvement even though only a potentially
suboptimal power allocation scheme is used. We also studied the secure
relaying problem when the eavesdropper's CSI is unknown. Instead of
maximizing the secrecy rate, a more reasonable relaying scheme with
both PCJ and FCJ is proposed in which a target QoS for the relay
network is met, and only the remaining resources are used for
jamming. These schemes are shown to provide large gains in terms of
the difference in the mutual information between the desired receiver
and the eavesdropper.  In particular, FCJ is shown to be a better
choice when the eavesdropper's CSI is unavailable since the ability to
exploit additional jamming subspace dimensions is preferable when the
transmitters possess no information about the eavesdropper.


\appendices
\section{SDP for problem (\ref{eq:qzb})} \label{sec:appsdp}
Let $\tilde{\mathbf{Q}}_{zb^\prime} = \mathbf{T}_b^\prime \mathbf{Q}_{zb^\prime}
\mathbf{T}_b^{\prime H}$, where $\mathbf{T}_b^\prime$ is normalized such that $\textrm{tr}(\tilde{\mathbf{Q}}_{zb^\prime})=\textrm{tr}(\mathbf{Q}_{zb^\prime})$.  Problem \eqref{eq:qzb} is equivalent to
\begin{subequations}
\begin{align}
\min_{\tilde{\mathbf{Q}}_{zb^\prime} \succeq 0, \mu \ge 0}& \quad \mu \\
\textrm{s.t.}& \quad \mathbf{t}_{a}^H \mathbf{H}_{ae}^H (\mathbf{H}_{be} \tilde{\mathbf{Q}}_{zb^\prime} \mathbf{H}_{be}^H + \sigma^2 \mathbf{I})^{-1} \mathbf{H}_{ae} \mathbf{t}_{a} \le \mu \label{eq:schurc}\\
						 & \quad \textrm{tr}(\tilde{\mathbf{Q}}_{zb^\prime}) \le p_b.
\end{align}
\end{subequations}
Using the Schur complement \cite{Golub_Matrix96}, constraint \eqref{eq:schurc} can be written as
\begin{equation} \label{eq:schurc1}
\left[\begin{array}{ll}
\mu & \mathbf{H}_{ae}^H \mathbf{t}_a^H \\
\mathbf{H}_{ae} \mathbf{t}_a & \mathbf{H}_{be} \tilde{\mathbf{Q}}_{zb^\prime} \mathbf{H}_{be}^H + \sigma^2 \mathbf{I}
\end{array} \right] \succeq 0.
\end{equation}
Combining \eqref{eq:schurc1} with the trace constraint and the semidefinite constraints on $\tilde{\mathbf{Q}}_{zb^\prime}$, the equivalent problem becomes
\begin{subequations} \label{eq:sdp}
\begin{align}
\min& \quad \mu \\
\textrm{s.t.}& \quad \textrm{tr}(\tilde{\mathbf{Q}}_{zb^\prime}) \le p_b,  \tilde{\mathbf{Q}}_{zb^\prime} \succeq 0, \mu \ge 0, \mathbf{h}_{br} \tilde{\mathbf{Q}}_{zb^\prime} = \mathbf{0}\\
						 & \quad
\left[\begin{array}{ll}
\mu & \mathbf{H}_{ae}^H \mathbf{t}_a^H \\
\mathbf{H}_{ae} \mathbf{t}_a & \mathbf{H}_{be} \tilde{\mathbf{Q}}_{zb^\prime} \mathbf{H}_{be}^H + \sigma^2 \mathbf{I}
\end{array} \right] \succeq 0 .
\end{align}
\end{subequations}
This is an SDP that consists of a linear objective function, a linear equality constraint, and a set of linear matrix inequalities (LMIs) \cite{Boyd_Convex04}, and thus can be solved efficiently, and $\mathbf{T}_b^\prime$ can be obtained via the eigenvalue decomposition of $\tilde{\mathbf{Q}}_{zb^\prime}$.

\section{Proof of Lemma \ref{rk1}} \label{sec:apprk1}
Given any jamming beamformer $\mathbf{T}_b^\prime$, \eqref{eq:qzb} becomes
\begin{align} \label{eq:qzb1}
\min_{\mathbf{Q}_{zb^\prime} \succeq 0} \  f(\mathbf{Q}_{zb^\prime}) \qquad \textrm{s.t.} \  \textrm{tr}(\mathbf{Q}_{zb^\prime}) \le p_b
\end{align}
where $f(\mathbf{Q}_{zb^\prime})=\mathbf{t}_{a}^H \mathbf{H}_{ae}^H (\mathbf{H}_{be}
\mathbf{T}_b^\prime \mathbf{Q}_{zb^\prime}
\mathbf{T}_b^{\prime H} \mathbf{H}_{be}^H + \sigma^2 \mathbf{I})^{-1} \mathbf{H}_{ae} \mathbf{t}_{a}$, and the Lagrangian of \eqref{eq:qzb1} is
\begin{equation}
L(\mathbf{Q}_{zb^\prime},\lambda,\mathbf{\Phi}) = f(\mathbf{Q}_{zb^\prime})
 + \lambda (\textrm{tr}(\mathbf{Q}_{zb^\prime})- p_b) - \textrm{tr}(\mathbf{\Phi} \mathbf{Q}_{zb^\prime})
\end{equation}
where $\lambda \ge 0$ is the Lagrange multiplier associated with the inequality constraint $\textrm{tr}(\mathbf{Q}_{zb^\prime}) \le p_b$, and $\mathbf{\Phi} \succeq 0$ is the Lagrange multiplier associated with the inequality constraint $\mathbf{Q}_{zb^\prime} \succeq 0$. Next, we can obtain the necessary conditions for the optimal $\mathbf{Q}_{zb^\prime}$ by using the Karush-Kuhn-Tucker (KKT) conditions:
\begin{align}
&\textrm{tr}(\mathbf{Q}_{zb^\prime}) \le  p_b, \ \mathbf{Q}_{zb^\prime} \succeq 0, \
						  \lambda \ge 0, \ \mathbf{\Phi} \succeq 0 \\
						  &\textrm{tr}(\mathbf{\Phi} \mathbf{Q}_{zb^\prime}) = 0 \label{trqphi} \\
						  &\lambda (\textrm{tr}(\mathbf{Q}_{zb^\prime})- p_b) = 0\\
						  &\mathbf{\Phi} - \mathbf{\Theta} = \lambda \mathbf{I} \label{thetarank}
\end{align}
where $\mathbf{\Theta}$ is obtained by differentiating $f(\mathbf{Q}_{zb^\prime})$ with respect to $\mathbf{Q}_{zb^\prime}$,
and is given by
\begin{equation*}
\mathbf{\Theta} = - \mathbf{H}_{be}^H
\mathbf{T}_b^{\prime H} (\mathbf{H}_{be}
\mathbf{T}_b^\prime \mathbf{Q}_{zb^\prime}
\mathbf{T}_b^{\prime H} \mathbf{H}_{be}^H + \sigma^2 \mathbf{I})^{-1}
\mathbf{H}_{ae} \mathbf{t}_{a} \mathbf{t}_{a}^H \mathbf{H}_{ae}^H
(\mathbf{H}_{be}
\mathbf{T}_b^\prime \mathbf{Q}_{zb^\prime}
\mathbf{T}_b^{\prime H} \mathbf{H}_{be}^H + \sigma^2 \mathbf{I})^{-1}
\mathbf{T}_b^\prime \mathbf{H}_{be} .
\end{equation*}
Since $\mathbf{t}_a$ is a vector, it is obvious that $\mathbf{\Theta}$ is a rank-one negative semidefinite matrix.

For the case that $\lambda = 0$, according to \eqref{thetarank}, we
have $\mathbf{\Theta} = \mathbf{\Phi}$. Since $\mathbf{\Theta}$ has a
negative eigenvalue, $\mathbf{\Phi}$ will also have a negative
eigenvalue, which contradicts the fact that $\mathbf{\Phi}$ is
positive semidefinite. Thus $\lambda$ can only be positive. For
$\lambda > 0$, according to \eqref{thetarank}, we know that
$\mathbf{\Phi}-\mathbf{\Theta}$ is a positive definite
matrix. Therefore, $\mathbf{\Phi}$ has at least $N-1$ positive
eigenvalues, \textit{i.e.} $rank(\mathbf{\Phi}) \ge N-1$, in order to
keep $\mathbf{\Phi} - \mathbf{\Theta} \succ 0$.

Assuming
$\lambda_i(\mathbf{\mathbf{\Phi}})$ and
$\lambda_{i}(\mathbf{Q}_{zb^\prime}), i=\{1,2,\dots,N\}$ are eigenvalues of
$\mathbf{\mathbf{\Phi}}$ and $\mathbf{Q}_{zb^\prime}$, respectively, in non-increasing order,
and due to the fact that $\mathbf{\mathbf{\Phi}}$ and $\mathbf{Q}_{zb^\prime}$ are both
positive semidefinite matrices, we have  $ \textrm{tr}(\mathbf{\mathbf{\Phi}}
\mathbf{Q}_{zb^\prime}) \ge \sum_{i=1}^{N}
\lambda_i(\mathbf{\mathbf{\Phi}})
\lambda_{N-i+1}(\mathbf{Q}_{zb^\prime})$.  Combining this observation
with \eqref{trqphi}, we also have 
\begin{equation}
\sum_{i=1}^{N} \lambda_i(\mathbf{\mathbf{\Phi}}) \lambda_{N-i+1}(\mathbf{Q}_{zb^\prime}) = 0. \label{eq:phiq0}
\end{equation}
Thus we can conclude that $rank(\mathbf{\Phi}) \neq N$, since otherwise all eigenvalues of $\mathbf{Q}_{zb^\prime}$ are zero and no jamming signals are transmitted. Combining this conclusion and the observation that $rank(\mathbf{\Phi}) \ge N-1$, we can conclude that $rank(\mathbf{\Phi}) = N-1$. Therefore, according to \eqref{eq:phiq0}, we have $\lambda_1(\mathbf{Q}_{zb^\prime})>0$ and $\lambda_{i \neq 1}(\mathbf{Q}_{zb^\prime})=0$, which indicates that $rank(\mathbf{Q}_{zb^\prime}) = 1$, and the proof is completed.

\section{Proof of Proposition \ref{simplegsvd}} \label{sec:appspgsvd}
According to the signal model given in Section \ref{sec:repcod}, the signals received by Eve during both phases can be combined together as
\begin{equation}
\mathbf{y}_e=
\left[ \begin{array}{l}
\mathbf{H}_{ae} \mathbf{T}_{a} \mathbf{D}_{a}  \\
 \mathbf{H}_{re} \mathbf{T}_{r} \mathbf{D}_{r}
\end{array} \right]
\mathbf{z}
+
\left[ \begin{array}{l}
\mathbf{n}_{e1}  \\
 \mathbf{n}_{e2}
\end{array} \right]
=\widetilde{\mathbf{H}}_{e} \mathbf{z} + \widetilde{\mathbf{n}}_e
\end{equation}
where $\mathbf{D}_k = diag\{\sqrt{p_{k,i}}\}$ and $\mathbb{E}(\mathbf{z} \mathbf{z}^H)=\mathbf{Q}_{z}=\mathbf{I}$.

Using the transmit beamformers in \eqref{eq:tbfgsvd}, and denoting $\mathbb{E}(\mathbf{z}_a \mathbf{z}_a^H)=\mathbf{Q}_{za}=diag\{p_{a,1},\cdots,p_{a,s}\}$, $\mathbb{E}(\mathbf{z}_r \mathbf{z}_r^H)=\mathbf{Q}_{zr}=diag\{p_{r,1},\cdots,p_{r,s}\}$,
the mutual information between Alice and Bob is
\begin{equation}
I_d=\min \left\{ \frac{1}{2} \log_2 \det (\mathbf{I}+\frac{1}{\sigma^2} \mathbf{H}_{ar} \mathbf{T}_a \mathbf{Q}_{za} \mathbf{T}_a^H \mathbf{H}_{ar}^H),\frac{1}{2} \log_2 \det (\mathbf{I}+\frac{1}{\sigma^2} \mathbf{H}_{rb} \mathbf{T}_r \mathbf{Q}_{zr} \mathbf{T}_r^H \mathbf{H}_{rb}^H) \right\}
\end{equation}
where
\begin{align}
\frac{1}{2} \log_2 \det (\mathbf{I}+\frac{1}{\sigma^2} \mathbf{H}_{ar} \mathbf{T}_a \mathbf{Q}_{za} \mathbf{T}_a^H \mathbf{H}_{ar}^H)
					 &= \frac{1}{2} \log_2 \det (\mathbf{I}+\frac{1}{\sigma^2} \mathbf{S}_{ar} \mathbf{Q}_{za} \mathbf{S}_{ar}^H) \notag \\	
					 &= \frac{1}{2} \log_2 \prod_{i=1}^s \left( 1 + \frac{p_{a,i} s_{ar,i}^2}{\sigma^2 ||\mathbf{R}_a^{-1}||^2} \right) \\
\frac{1}{2} \log_2 \det (\mathbf{I}+\frac{1}{\sigma^2} \mathbf{H}_{rb} \mathbf{T}_r \mathbf{Q}_{zr}  \mathbf{T}_r^H \mathbf{H}_{rb}^H)
					&= \frac{1}{2} \log_2 \det (\mathbf{I}+\frac{1}{\sigma^2} \mathbf{S}_{rb} \mathbf{Q}_{zr} 	\mathbf{S}_{rb}^H) \notag \\	
					 &= \frac{1}{2} \log_2 \prod_{i=1}^s \left( 1 + \frac{p_{r,i} s_{rb,i}^2}{\sigma^2 ||\mathbf{R}_r^{-1}||^2} \right).			
\end{align}
For Eve, we have
\begin{equation}
I_e=\min \left \{ \frac{1}{2} \log_2 \det (\mathbf{I}+\frac{1}{\sigma^2} \widetilde{\mathbf{H}}_{ar} \mathbf{Q}_{za} \widetilde{\mathbf{H}}_{ar}^H),\frac{1}{2} \log_2 \det (\mathbf{I}+\frac{1}{\sigma^2} \widetilde{\mathbf{H}}_e \mathbf{Q}_{z} \widetilde{\mathbf{H}}_e^H) \right \}
\end{equation}
where
\begin{align}
\frac{1}{2} \log_2 \det (\mathbf{I}+\frac{1}{\sigma^2} \widetilde{\mathbf{H}}_e \mathbf{Q}_{z} \widetilde{\mathbf{H}}_e^H)
								&= \frac{1}{2} \log_2 \det (\mathbf{I}+\frac{1}{\sigma^2} \widetilde{\mathbf{H}}_e^H \widetilde{\mathbf{H}}_e) \notag \\
					     &= \frac{1}{2} \log_2 \det (\mathbf{I}+\frac{1}{\sigma^2} (\mathbf{D}_a^H \mathbf{S}_{ae}^H \mathbf{S}_{ae} \mathbf{D}_a +
					     \mathbf{D}_r^H \mathbf{S}_{re}^H \mathbf{S}_{re} \mathbf{D}_r) ) \notag \\
					     &= \frac{1}{2} \log_2 \prod_{i=1}^s \left( 1 + \frac{p_{a,i} s_{ae,i}^2}{\sigma^2 ||\mathbf{R}_a^{-1}||^2}+
					     \frac{p_{r,i} s_{re,i}^2}{\sigma^2 ||\mathbf{R}_r^{-1}||^2} \right),
\end{align}
and according to the same secrecy constraints in \eqref{eq:seccons},
the secrecy rate \eqref{eq:cap_gsvd} can be obtained.


\bibliographystyle{IEEEtran}
\bibliography{IEEEabrv,mybibfile}

\end{document}